\documentclass[a4paper,fleqn,usenatbib]{mnras}
\usepackage[T1]{fontenc}
\usepackage{ae,aecompl}
\usepackage{graphicx}
\graphicspath{{images_de_larticle/}}
\usepackage{amssymb}
\usepackage{amsmath}
\usepackage[usenames,dvipsnames]{xcolor}
\definecolor{notecolor}{rgb}{0.8,0,0}

\usepackage{siunitx}
\def\lya{Ly$\alpha$~} 

\title[The self-shielding of the IGM at high redshift]{
Self-shielding of hydrogen in the IGM during the epoch of reionization
}
\author[Chardin et al.]{Jonathan Chardin$^{1}$\thanks{E-mail: jc@ast.cam.ac.uk}, 
Girish Kulkarni$^{1}$ and Martin G. Haehnelt$^{1}$\\
$^1$Institute of Astronomy and Kavli Institute of Cosmology,
University of Cambridge, Madingley Road, Cambridge CB3 0HA, UK}
\date{Accepted ---. Received ---; in original form ---}

\pubyear{2017}

\begin{document}
\label{firstpage} 
\pagerange{\pageref{firstpage}--\pageref{lastpage}}
\maketitle

\begin{abstract}
 We investigate self-shielding of intergalactic hydrogen against
 ionizing radiation in radiative transfer simulations of cosmic
 reionization carefully calibrated with \lya forest data.  While
 self-shielded regions manifest as Lyman-limit systems in the
 post-reionization Universe, here we focus on their evolution during
 reionization (redshifts $z=6$--$10$).  At these redshifts, the
 spatial distribution of hydrogen-ionizing radiation is highly
 inhomogeneous, and some regions of the Universe are still neutral.
 After masking the neutral regions and ionizing sources in the
 simulation, we find that the hydrogen photoionization rate depends on
 the local hydrogen density in a manner very similar to that in the
 post-reionization Universe.  The characteristic physical hydrogen
 density above which self-shielding becomes important at these
 redshifts is about $n_\mathrm{H}\sim 3\times 10^{-3}$~cm$^{-3}$, or
 $\sim 20$ times the mean hydrogen density, reflecting the fact that
 during reionization photoionization rates are typically low enough
 that the filaments in the cosmic web are often self-shielded.  The
 value of the typical self-shielding density decreases by a factor of
 3 between redshifts $z=3$ and $10$, and follows the evolution of the
 average photoionization rate in ionized regions in a simple fashion.
 We provide a simple parameterization of the photoionization rate as a
 function of density in self-shielded regions during the epoch of
 reionization.
\end{abstract}


\section{Introduction}
\label{intro}

The epoch of hydrogen reionization is thought to have begun with the
formation of the first sources of radiation in high-density peaks of
the collapsing large-scale structure in the Universe
\citep{2000ApJ...535..530G, 2001PhR...349..125B, 2005MNRAS.361..577C}.
Escaping Lyman continuum photons from these sources create growing
H~\textsc{ii} regions around them, which grow and merge with
H~\textsc{ii} regions associated with other ionizing sources
\citep{1987ApJ...321L.107S, 2016arXiv160605352A, 2016MNRAS.462..250M,
  2017MNRAS.468.2176S}.  Reionization is said to be complete when
these ionized regions percolate, fill the cosmic volume, and establish
a metagalactic UV background \citep{2000ApJ...530....1M}.  Current
constraints from \lya and CMB data suggest that hydrogen reionization
was complete at redshift $z\gtrsim 6$ \citep{2015ApJ...802L..19R,
  2015MNRAS.454L..76M, 2016A&A...596A.108P, 2017arXiv170203687B}.

Although the Universe is highly ionized at redshifts $z<6$, the
densest regions of the cosmic web remain neutral due to high
recombination rates \citep{2000ApJ...530....1M, 2009MNRAS.394..960C}.
These neutral islands span a range of different neutral hydrogen
column densities and are said to have self-shielded against
hydrogen-ionizing radiation \citep{2013MNRAS.430.2427R}.
Self-shielding is increasingly effective as the column density
$N_\mathrm{HI}$ exceeds $\sim 1/\sigma\approx 10^{17}~\mathrm{cm}^2$,
where $\sigma$ is the frequency-dependent ionization
cross-section of hydrogen.  In the post-reionization Universe, such
systems are observed as Lyman-limit systems in the spectra of distant
sources such as quasars \citep[e.g.,][]{2014MNRAS.445.1745W}.  The
abundance of these self-shielded systems depends on the cosmological
gas density distribution and the amplitude of the relatively uniform
hydrogen photoionizing background.  Modern cosmological simulations
successfully reproduce the observed abundance and column density
distribution of these self-shielded systems at redshifts $z=2$--$5$ in
the post-reionization Universe \citep{2011ApJ...743...82M,
  2013MNRAS.436.2689A}.

On the other hand, properties of self-shielded regions before the
completion of reionization, at redshifts $z\gtrsim 6$, are less
well-understood.  Self-shielded regions at these redshifts represent
the residual neutral hydrogen content in H~\textsc{ii} regions.  But
H~\textsc{ii} regions associated with individual sources of ionizing
radiation have still not overlapped at these redshifts.  As a result,
the hydrogen photoionization rate has large spatial fluctuations
\citep{2015MNRAS.453.2943C}.  Under such conditions, simulating the
abundance and other properties of self-shielded regions requires an
accurate treatment of reionizing sources and the photoionization rate
fluctuations \citep{2005MNRAS.363.1031F, 2013ApJ...771...35K,
  2014ApJ...787..146K}.  These high-redshift Lyman-limit systems are
important as they can affect the interpretation of observations such
as the rapid decrease in the abundance of Ly$\alpha$ emitters
\citep{2009MNRAS.394..960C, 2013MNRAS.429.1695B, 2015MNRAS.452..261C,
  2015MNRAS.446..566M, 2015MNRAS.454..681K}.  As a reservoir of
neutral gas, they can also affect the 21~cm signal from the epoch of
reionization, which is being targetted by many ongoing and upcoming
experiments \citep{2014MNRAS.440.1662S, 2015MNRAS.449.3202W,
  2016MNRAS.458..135S, 2016MNRAS.463.2583K}.

In this paper, we present high-resolution radiation hydrodynamics
cosmological simulations of the intergalactic medium during the epoch
of reionization, and characterize the density scale at which
self-shielding occurs under a range of assumptions regarding the
ionizing sources.  These simulations are calibrated to reionization
constraints from CMB and Ly$\alpha$ data following the procedure
adopted by \citet{2015MNRAS.453.2943C}.  The simulations presented by
\citet{2015MNRAS.453.2943C} were able to reproduce the evolution of
the UV luminosity function of galaxies during and after reionization
while at the same time reproducing a range of IGM properties inferred
from observations of the \lya forest in the post-reionization phase.
We use similar simulations in the present study to investigate the
evolution of the self-shielding properties of hydrogen during and
after reionization.  We describe our simulations, and give details of
the reionization histories considered in this paper in
Section~\ref{Simu}.  We discuss self-shielding in our simulations in
Section~\ref{SS_simualtions_after_overlap}.  In
Section~\ref{discussion}, we test our results for numerical
convergence and dependence on ionizing source properties, and
summarise our conclusions in Section~\ref{prospects}.  Our
$\Lambda$CDM cosmological model has $\Omega_\mathrm{b}=0.048$,
$\Omega_\mathrm{m}=0.3175$, $\Omega_\Lambda=0.6825$, $h=0.6711$,
$n=0.9624$, $\sigma_8=0.8344$, and $Y_\mathrm{He}=0.24$
\citep{2014A&A...571A..16P}.

\section{Simulations}
\label{Simu}

The simulations presented in this paper are performed in two steps.
Hydrodynamical cosmological simulations are run as a first step, as
described in more detail by \citet{2015MNRAS.453.2943C,
  2017MNRAS.465.3429C}.  As the second step, sources of
hydrogen-ionizing radiation are set up in the simulation box, and
radiative transfer calculations are performed.

The cosmological simulations of the evolution of the dark matter and
gas hydrodynamics were performed using the \textsc{ramses} code
\citep{2002A&A...385..337T}.  The simulation evolves $512^3$ dark
matter particles.  Gas hydrodynamics is evolved on a coarse, fixed
grid discretized in $512^3$ cells.  We did not employ the adaptive
mesh refinement option of \textsc{ramses}.  Although we later
post-process this cosmological simulation for radiative transfer, we
also implement a uniform UV background model in the simulation to get
a realistic gas density structure.  We use the metagalactic UV
background model of \citet[][HM12 hereafter]{2012ApJ...746..125H}, in
which the time evolution of the space-averaged UV background intensity
is calculated by solving a global radiative transfer equation with an
empirical model opacity due to hydrogen and helium, and a source
function based on the observed UV luminosity function of quasars and
star-forming galaxies.  In the HM12 model, the gas is heated to
$T_0\sim 10^4$~K at redshift $z=15$, where $T_0$ is the temperature at
the mean gas density.  The subsequent thermal evolution is calculated
in \textsc{ramses} assuming photoionization equilibrium.  As a result
gas cools down to $T_0\sim 7\times 10^3$ K at $z\sim 6$ and then
undergoes another episode of heating at $z\sim 2$--$4$ corresponding
to helium reionization \citep{2015arXiv1410.1531P,
  2015MNRAS.453.2943C}.  This has a corresponding effect on the
structure of gas density due to pressure smoothing
\citep{2017arXiv170408366R, 2009MNRAS.394.1812P}.  Gas density
snapshots were taken from redshift $z=100$ to $z=2$ at 40~Myr
intervals.

\subsection{Radiative transfer} 
\label{postRT}

The radiative transfer calculations were performed in post-processing
with \textsc{aton} (\citealt{2008MNRAS.387..295A}).  \textsc{aton} is
a \textsc{gpu}-enabled radiative transfer code that implements a
moment-based description of the radiative transfer equation.  We
employ a monochromatic radiation field, which assumes that all
ionizing photons have an energy of 20.27~eV.  This value corresponds
to the mean energy of a $5\times 10^4$~K blackbody spectrum, which is
a close approximation to the spectral energy distribution of a stellar
population with a Salpeter initial mass function with stellar masses
in the range 1--100~M$_\odot$ \citep{2009A&A...495..389B}.

Haloes are identified in the base simulation using the \textsc{hop}
halo finder \citep{1998ApJ...498..137E}, with minimum halo mass
consisting of 10 dark matter particles.  Radiative transfer is started
from the redshift at which the first halo collapses.  In our fiducial
simulation, ionizing sources are placed in dark matter haloes and
assumed to emit continuously.  As we discuss below, we also consider
simulations in which ionizing sources are placed on the sides of the
simulation box.  The luminosities of ionizing sources are set assuming
a linear scaling of the luminosity with halo mass
\citep{2006MNRAS.369.1625I, 2012A&A...548A...9C, 2014A&A...568A..52C}.
The normalization of this luminosity-mass relation is assumed to vary
with redshift and is chosen so that the integrated comoving ionizing
emissivity is similar to that of the HM12 uniform UV background model,
but somewhat modified so as to obtain an improved match with the
hydrogen photoionization rates inferred from \lya forest data
\citep{2015MNRAS.453.2943C}.

We self-consistently follow recombination radiation and do not use the
on-the-spot approximation \citep{2013MNRAS.430.2427R}: atoms recombine
and radiate in an isotropic fashion, and are not directly reabsorbed
by the freshly recombined hydrogen.  Note that the impact of
recombination radiation is likely to be underestimated in our
simulations.  Recombination radiation is expected to be emitted at a
frequency close to the ionization threshold, where the hydrogen
ionization cross-section $\sigma=6.3\times 10^{18}~\mathrm{cm}^2$ is
the highest, while our monochromatic frequency-averaged method tends
to underestimate its absorption probability by assigning
$\sigma=1.6\times 10^{18}~\mathrm{cm}^2$.  The mean free path of these
photons is thus larger than it should be, and a more accurate
multi-frequency description would reduce the impact of recombinations.
Moreover, we do not take into account the redshifting of the
recombination photons by peculiar velocities of the emitters or the
Hubble flow.  In reality, recombination photons cannot travel to large
cosmological distances without being redshifted to frequencies below
the Lyman limit.  Therefore, neglecting the cosmological redshifting
of recombination radiation could in principle result in an
overestimation of the photoionization rate on large scales.  However,
our simulation box is small enough that this is not a significant
concern.  The mean free path of Redshifting is potentially a
  concern when the mean free path of photons exceeds the box size.
  This does not happen in our simulation until $z\lesssim 5$.  Then,
  at these low redshifts, redshifting is not a concern because a
  uniform UV background is established.

\begin{table}
  \begin{center}
    \begin{tabular}{lccl}
      \hline
      Model & Box size & $N_\mathrm{grid}$ & Source model \\
      & (cMpc$/h$) & & \\
      \hline
      L20N512 & 20 & $512^3$ & Haloes \\
      L20N512pp & 20 & $512^3$ & Plane-parallel \\
      L20N512fs & 20 & $512^3$ & Restricted haloes \\
      L10N512pp & 10 & $512^3$ & Plane-parallel \\
      L10N512fs & 10 & $512^3$ & Restricted haloes \\
      \hline 
    \end{tabular}
  \end{center}
  \caption{Simulations used in this work: L20N512 is our fiducial
    simulation.  The three source models are described in
    the text.}
  \label{tab:models}
\end{table}

\subsection{Ionizing sources} 
\label{set_model}

\begin{figure}
   \begin{center}
      \includegraphics[width=\columnwidth]{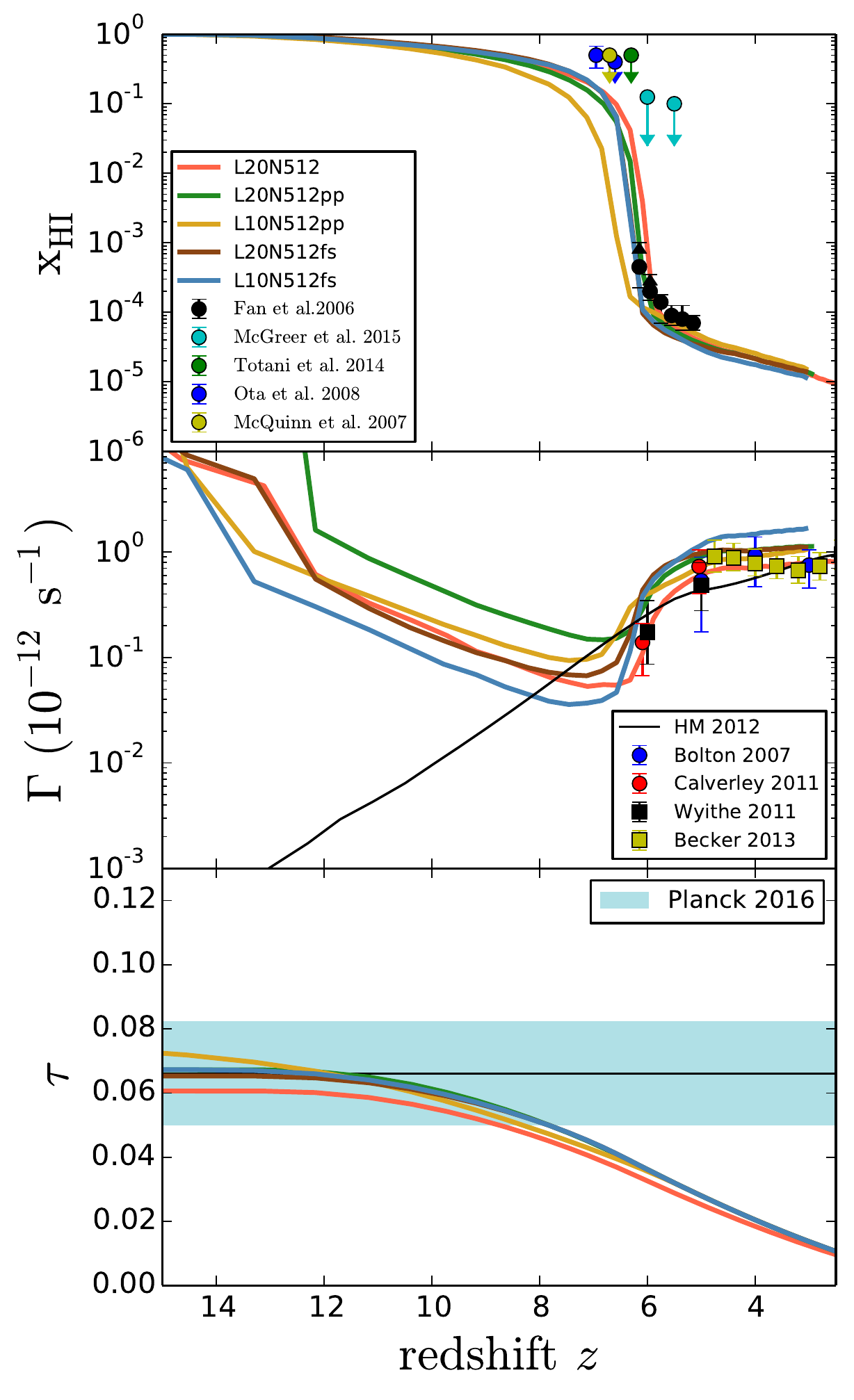}   
  \caption{Calibration of the five simulations listed in
    Table~\ref{tab:models}.  The top panel shows the evolution of the
    average neutral hydrogen fraction.  Measurements shown are from
    \citet{2006AJ....132..117F}, \citet{2015MNRAS.447..499M},
    \citet{2014PASJ...66...63T}, \citet{2008ApJ...677...12O}, and
    \citet{2007MNRAS.381...75M}.  The middle panel shows the evolution
    of the average photoionization rate compared with measurements by
    \citet{2007MNRAS.382..325B}, \citet{2011MNRAS.412.2543C},
    \citet{2011MNRAS.412.1926W} and \citet{2013MNRAS.436.1023B}.  The
    bottom panel shows the Thompson scattering optical depth, compared
    with the measurement from \citet{2016A&A...596A.108P}.}
    \label{xion_HI_vs_z}
  \end{center}
 \end{figure}

Our fiducial simulation is performed in a cubical box 20 comoving
Mpc$/h$ on a side with periodic boundary conditions.  In this model,
we place sources of ionizing radiation at the locations of dark matter
haloes, with a linear scaling of their luminosity with halo mass.  The
amplitude of this luminosity-mass relation is fixed by the HM12 model,
as described above.  For comparison, we also consider an equivalent
simulation box in which plane-parallel radiation fronts are propagated
in the three spatial directions \citep[cf.][]{2013MNRAS.430.2427R}.
This is achieved by placing radiation sources on the sides of the
cubical box.  We use 128 sources on each side.  These sources have
identical luminosities, which are chosen so that the total emissivity
agrees with that in the HM12 model.  A comparison of this
plane-parallel model with our fiducial model allows us to isolate the
effect of fluctuations in the photoionization rate due to sources.  To
check our results for numerical convergence, we also consider another
simulation with a box size of 10~cMpc$/h$.  The radiative transfer in
our simulations is also performed in two ways: with local sources
placed in dark matter haloes, and with plane-parallel radiation
fronts.

The spatial resolution of our fiducial simulation is 39.1~ckpc$/h$,
while the spatial resolution in the 10~cMpc$/h$ box is 19~ckpc$/h$.
At redshift $z=10$, the minimum and maximum halo masses in our
fiducial simulation are $M_\mathrm{min}=2\times 10^8$~M$_\odot$ and
$M_\mathrm{max}=5\times 10^{10}$~M$_\odot$.  In the 10~cMpc$/h$ box,
the minimum and maximum halo masses at $z=10$ are $3\times
10^{7}$~M$_\odot$ and $10^{10}$~M$_\odot$, respectively.  However,
when comparing the results from these simulations with our fiducial
model in the case of local sources, we restrict the range of halo
masses so that they are resolved by both boxes.  This range is chosen
to be between $M_\mathrm{min}=2\times 10^8$~M$_\odot$ and $5\times
10^{9}$~M$_\odot$.  Haloes with mass outside this range do not host
sources, even if present in the simulation box.

We consider five simulations as listed in Table~\ref{tab:models}, in
which the fiducial simulation is labelled as L20N512. Simulations with
the plane-parallel source model have labels with the suffix `pp', and
simulations in which sources occupy a restricted range of halo masses
are identified with labels that end in `fs'.

\begin{figure*}
   \begin{center}
      \includegraphics[width=\textwidth]{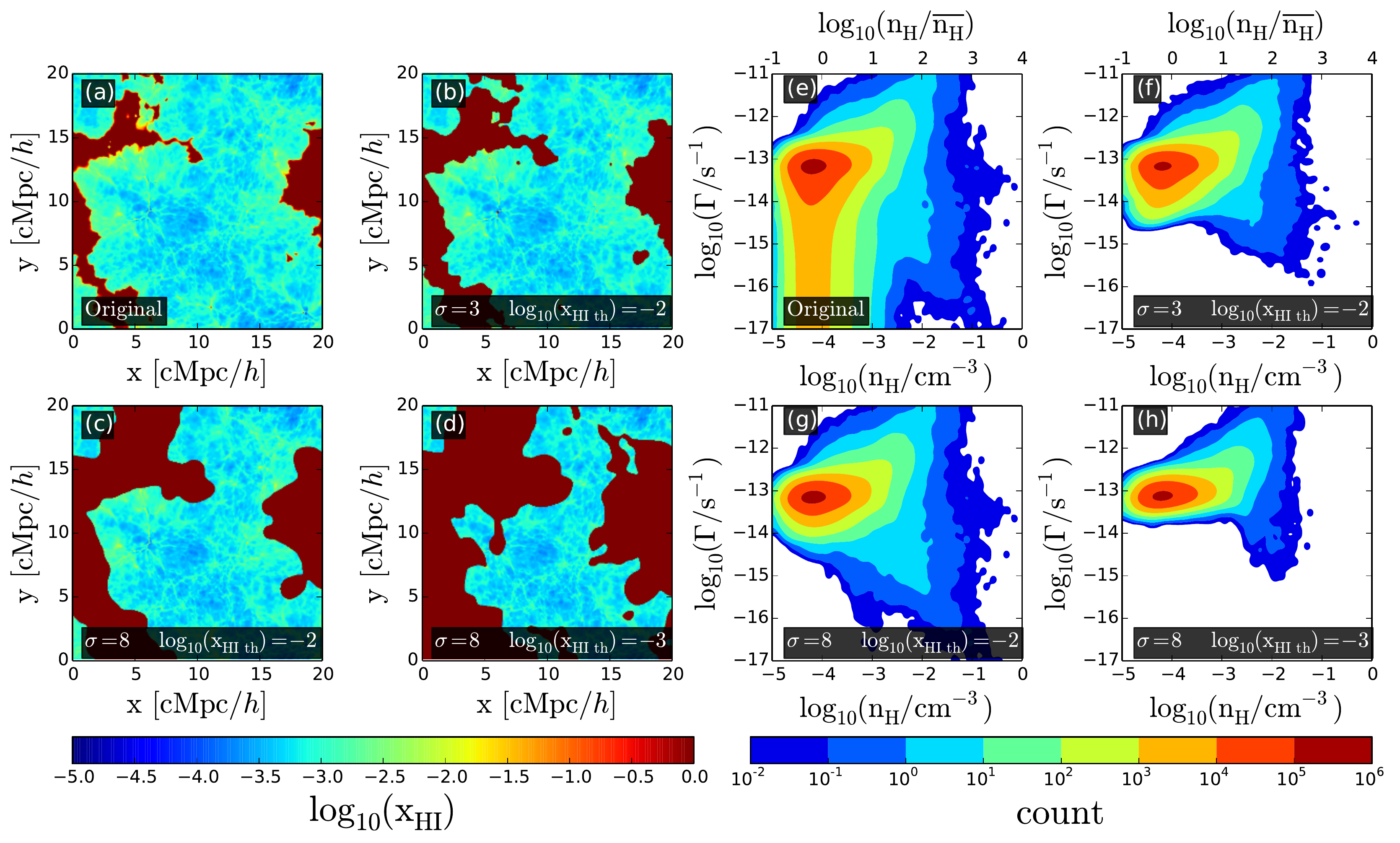}   
      \caption{Isolation of self-shielded regions in our simulations.
        The four panels on the left show the neutral hydrogen
        distribution at $z=7$ in our fiducial L20N512 simulation.  The
        four panels on the right show the corresponding
        photoionization rate density distribution.  The top left panel
        in each set shows the distributions before isolating the
        self-shielded regions.  The other three panels show the result
        of convolving the spatial neutral hydrogen distribution with a
        Gaussian kernel of width $\sigma$ and then, after convolution,
        discarding cells with a neutral hydrogen fraction less than
        $x_\mathrm{HI,th}$.}
  \label{mosaique111}
  \end{center}
\end{figure*}

\subsection{Calibration}
\label{reionization_hist}

As mentioned above, we use an ionizing emissivity that is close to
that of the HM12 model in our simulations.  Figure~\ref{xion_HI_vs_z}
shows the result.  The top panel of Figure~\ref{xion_HI_vs_z} shows
the redshift evolution of the average neutral fraction in our
different models.  As all of our models are calibrated to the same
HM12 reionization history, only small variations are seen between the
different models.  All models lie close to the values of the neutral
fraction derived from the \lya forest data of
\citet{2006AJ....132..117F}.  This is also reflected in the bottom
panel of Figure~\ref{xion_HI_vs_z}, which shows the electron
scattering optical depth in the simulations.  It is worth noting that
simulations with the plane-parallel source model show an almost
exactly similar reionization history compared to simulations with the
fiducial source model.  Adopting a tuned ionizing emissivity evolution
with sources on the edges of the cosmological box can therefore mimic
the simulations assuming the realistic positions of the sources.  This
similarity will allow us to compare directly the impact of the source
location on the self-shielding properties of the gas both at high and
low redshift.

The middle panel of Figure~\ref{xion_HI_vs_z} shows the redshift
evolution of the average hydrogen photoionization rate in ionized
regions in our different models.  The photoionization rate,
$\mathrm{\langle\Gamma\rangle}$, is averaged over the already ionized
regions (i.e., regions with $\mathrm{x_{HII}\ge 0.5}$).  The evolution
of the photoionization rate is consistent with observations in the
post-overlap phase.  Thus, regardless of ionizing source model
considered, our fine tuning of the emissivity evolution with redshift
results in a nearly identical reionization history.  First, at high
redshifts, the average photoionization rate drops with time due to
geometric dilution, as the photons inside ionized regions are consumed
at the edges of ionized regions.  Towards the end of reionization, the
mean free path of ionizing photons rapidly increases when distinct
ionized regions overlap, concomitantly increasing the photoionization
rate.  We note here that the agreement between the different models is
not perfect, particularly at high redshifts.  However, the
post-reionization evolution of the average photoionization rate is
always in good agreement with the observationally inferred
photoionization rate at low redshifts.  We therefore expect to see
some variation between the different models at high redshifts.  This
may also lead to small differences in the characteristics of
self-shielded regions in these simulations.  We need to keep in mind
those differences in the reionization histories when we interpret our
results below.  Finally, the bottom panel of
  Figure~\ref{xion_HI_vs_z} shows that electron scattering optical
  depth in the simulations agrees with the Planck measurements
  \citep{2016A&A...596A.108P}. 

\section{The self-shielding of hydrogen in our simulations}
\label{SS_simualtions_after_overlap}

Panel~(a) of Figure~\ref{mosaique111} shows the distribution of the
neutral hydrogen fraction $x_\mathrm{HI}$ in a slice of thickness
39.1~ckpc$/h$ at $z=7$ in our fiducial simulation.  The neutral
hydrogen fraction is unity in regions of the simulation that are yet
to be reionized; in the ionized regions the neutral hydrogen fraction
is generally low.  However, in the ionized regions, $x_\mathrm{HI}$
varies with the total gas density between $10^{-5}$ and $0.1$.  This
variation is partly due to fluctuations in the photoionization rate
and partly due to self-shielding in high-density regions.

In order to study the neutral fraction in ionized regions, we need to
isolate these regions in our simulation.  This is difficult due to the
complex morphology of these regions.  A criterion that depends on the
neutral hydrogen fraction at a location is unable to distinguish
between a self-shielded cell within an ionized region and a cell in a
region that is not yet ionized (the ``neutral region'' for short).  In
this work we disentangle the neutral and ionized regions in our
simulation box by applying a neighbourhood criterion: a cell with high
neutral fraction is deemed to lie in the neutral region if its
neighbours also have a high neutral fraction.

The number of neighbours considered in this selection will obviously
affect the result.  We decide this by first selecting neighbours in a
spherical region around a cell weighed, for simplicity, by a Gaussian.
The width $\sigma$ of the Gaussian and the threshold in the neutral
hydrogen fraction, $x_\mathrm{HI,th}$, are then chosen by visually
comparing the resulting selection with the original slice from the
simulation.  Figure~\ref{mosaique111} shows results for a range of
values of $\sigma$ and $x_\mathrm{HI,th}$.  For a given value of
  $\sigma$, we consider if cells with neutral hydrogen fraction
  greater than $x_\mathrm{HI,th}$ form contiguous neutral regions as
  opposed to isolated neutral cells in otherwise ionized regions.  If
the width of the Gaussian is too small, self-shielded regions are also
marked as neutral.  This is seen in Figure~\ref{mosaique111} in
panel~(b), which has $\sigma=3$~cells.  If the width is too large then
parts of the neutral region are added to the ionized regions.  Similar
errors in identification occur if extreme values of $x_\mathrm{HI,th}$
are used, as seen in panels~(c) and (d).  For the simulation and
redshift shown in Figure~\ref{mosaique111}, we use $\sigma=8$ cells
and $x_\mathrm{HI,th}=10^{-2}$, corresponding to panel~(c) of
Figure~\ref{mosaique111}.

The four panels in the right half of Figure~\ref{mosaique111} show the
photoionization rate distribution corresponding to the four panels in
the left half.  The average photoionization rate is $\sim
10^{-13}$~s$^{-1}$, consistent with the HM12 model.  But, as seen in
panel~(e), the photoionization rate distribution in the simulation is
characterized by high values ($\gtrsim 10^{-12}~\mathrm{s}^{-1}$)
close to sources in high density regions.  The photoionization rate is
the lowest in low density regions as some of these are not ionized yet
($\Gamma_\mathrm{HI}=0$).  The effect of the selection procedure
described above is seen in the other panels of
Figure~\ref{mosaique111} (panels f, g, and h): selecting for ionized
regions only leaves us with cells that have high photoionization rate
at low densities.  As we remove the neutral regions from
Figure~\ref{mosaique111}, a second branch starts to appear in the high
density part of the plot.  This branch is distinguished by a low
photoionization rate.  It is these cells that show the effect of
self-shielding.  However, in order to characterise them better we need
to separate them from the high density, high photoionization rate
cells located near sources of ionizing radiation.

\begin{figure}
   \begin{center}
      \includegraphics[width=\columnwidth]{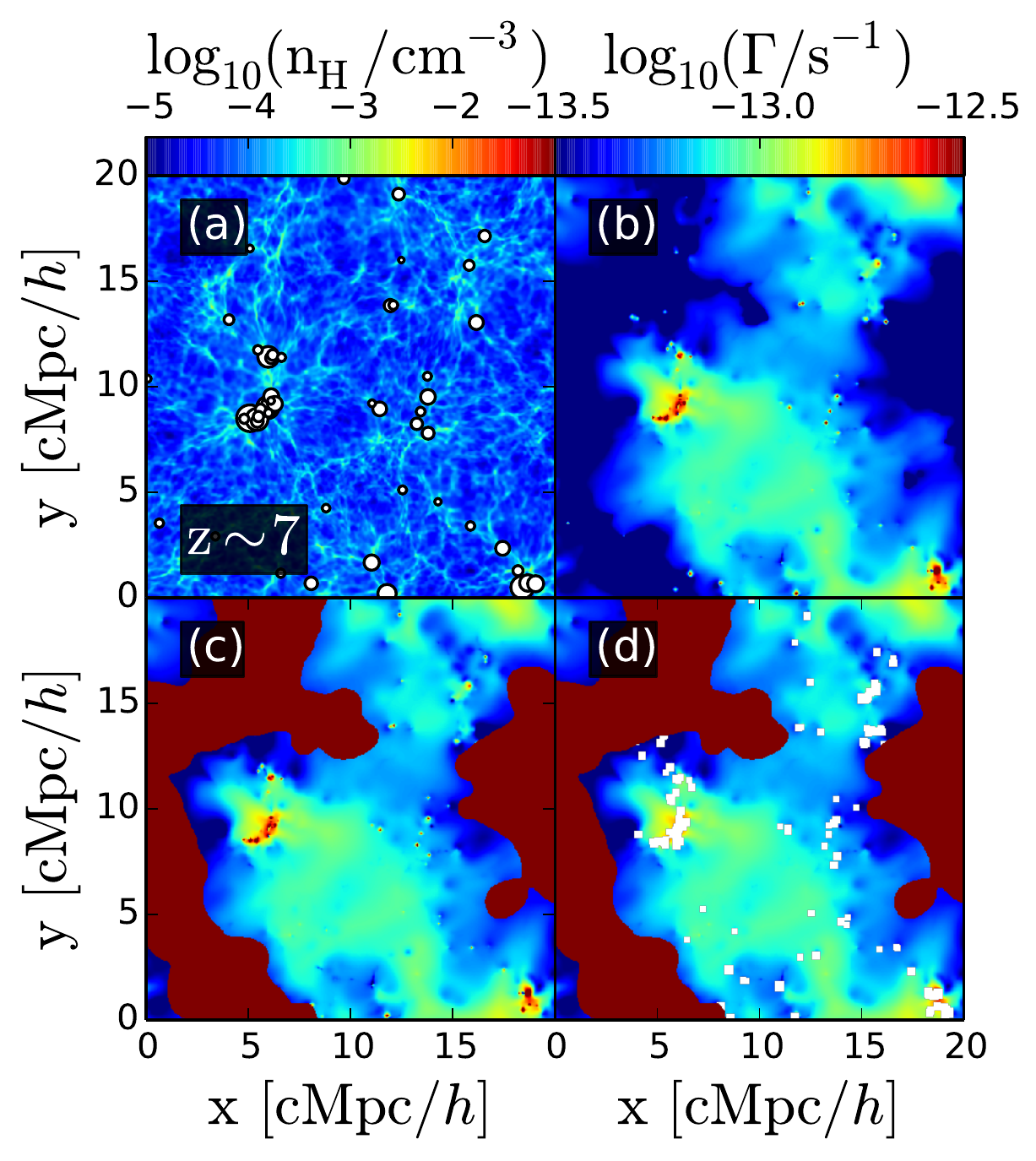}   
  \caption{Illustration of the source exclusion process in simulations
    with local source model.  Shown here is the fiducial L20N512
    simulation.  Panel~(a) shows a slice through the hydrogen number
    density distribution in the simulation at $z=7$.  Sources are
    marked with white circles, with the circle radii scaling as the
    source luminosity.  Panel~(b) shows the photoionization rate
    distribution corresponding to panel~(a).  Panel~(c) shows the
    photoionization rate distribution when the ionized regions are
    isolated by masking the neutral regions (brown) using the
    procedure described in Section~\ref{SS_simualtions_after_overlap}.
    Panel~(d) shows the masking of the sources.}
    \label{fig:source_rejection}
  \end{center}
\end{figure}

\begin{figure}
   \begin{center}
      \includegraphics[width=\columnwidth]{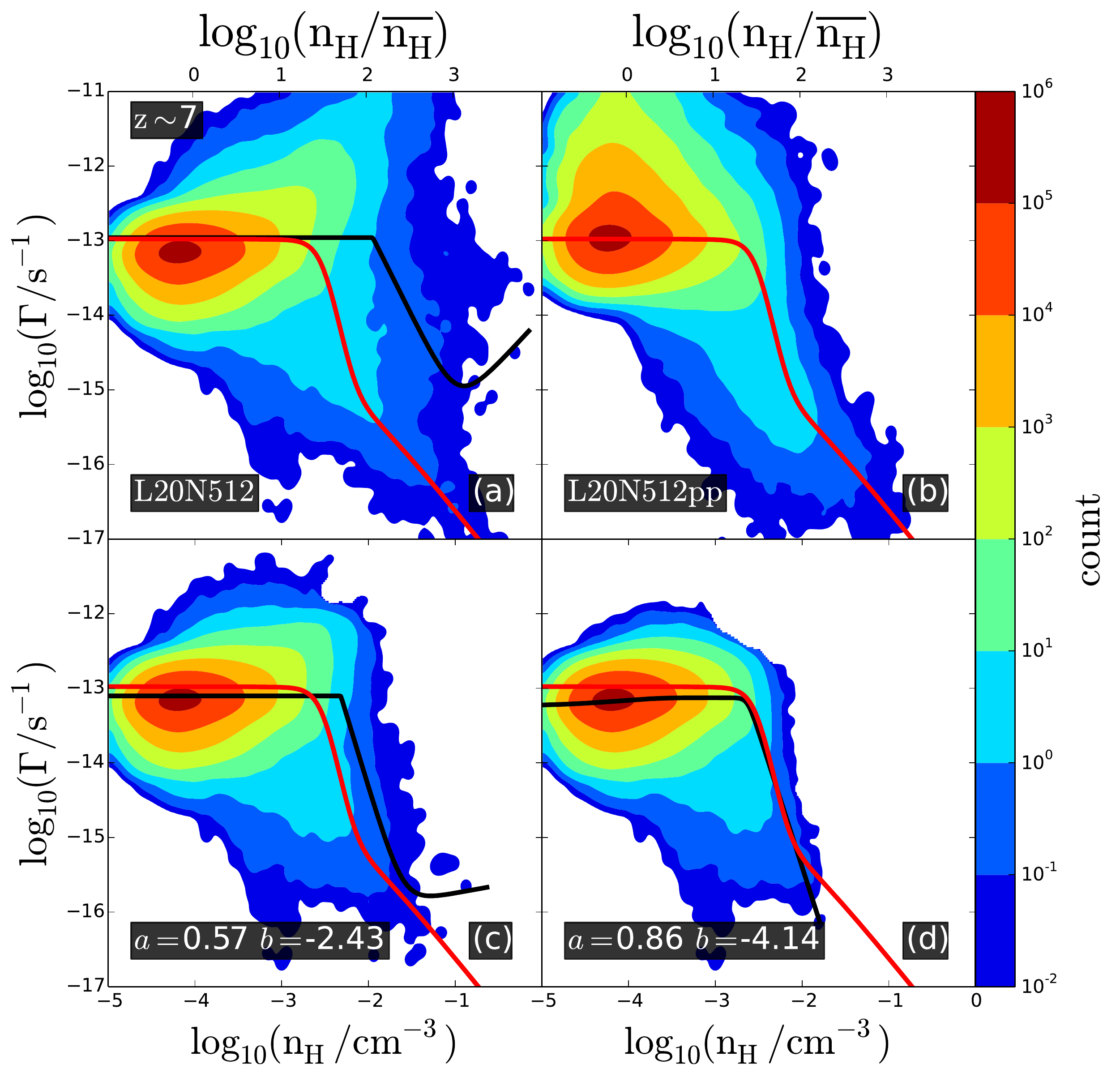}   
  \caption{Effect of source exclusion on the photoionization rate
    density distribution.  Panel~(a) shows the distribution from the
    fiducial L20N512 simulation at $z=7$.  Panel~(b) shows the
    photoionization rate density distribution from the L20N512pp
    simulation at the same redshift.  A comparison of these two panels
    shows that the top right part of the distribution in Panel~(a) is
    affected by enhanced photoionization rate near ionizing sources.
    Panels~(c) and (d) shows how this is corrected by masking these
    sources by following the procedure illustrated in
    Figure~\ref{fig:source_rejection}.  Cubical regions around source
    centres are masked, with the length of the cube taken to be
    $r=a\log_{10}(M_\mathrm{halo})+b$.  Panels~(c) and (d) show the
    result for two different values of the parameters $a$ and $b$.
    The red curve in panel~(b) shows the median photoionization rate;
    this curve is reproduced in panels~(c) and (d).  The black
      curves in all panels show the respective median photoionization
      rates.  We choose values of the parameters $a$ and $b$
    corresponding to panel~(d) so that the median photoionization rate
    agrees with that in the plane-parallel case.}
    \label{fig:source_rejection_gamma}
  \end{center}
 \end{figure}

In order to do this, we remove cells that contain sources from our
analysis. We also remove a number of cells that are adjacent to the
cells containing sources.  This procedure is illustrated in
Figure~\ref{fig:source_rejection}.  Panel~(a) of this figure shows the
distribution of total gas density in our fiducial simulation at $z=7$.
The white circles in this panel show the locations of ionizing
sources.  These correspond to the centres of masses of dark matter
haloes.  The areas of the circles are proportional to the source
luminosity, and thus the halo mass.  Panel~(b) of
Figure~\ref{fig:source_rejection} shows the corresponding slice for
the hydrogen photoionization rate.  A comparison of these panels shows
that the locations in the box with high photoionization rate
($\Gamma>10^{-12}~\mathrm{s}^{-1}$) correspond to the location of
sources.  Panel~(c) shows the effect of removing the regions that are
not reionized yet, following the procedure described above, while
Panel~(d) illustrates our masking of the high photoionization rate
cells.  We choose the number of cells to exclude based on the halo
luminosity and resolution.  The length of the cube excluded, centered
around the halo location, depends on the halo mass as
$r=a\log_{10}(M_\mathrm{halo})+b$.  In our fiducial simulation, a cell
size corresponds to the virial radius of a $3\times 10^9$~M$_\odot$
halo at $z=7$.  Excluding a few cells around larger haloes will ensure
that the high photoionization rate cells are completely removed from
our analysis.  We exclude a cubical region of about 8 to 14 cells
  on a side, depending on the halo mass.

The effect of removing sources on the photoionization rate
distribution is illustrated in Figure~\ref{fig:source_rejection_gamma}
for various values of $a$ and $b$.  This figure also shows how we
choose the values for $a$ and $b$.  Panel (a) of
Figure~\ref{fig:source_rejection_gamma} shows the photoionization rate
density distribution for our fiducial simulation at $z=7$, after
isolating the ionized regions.  We continue to see regions with high
density and high photoionization rate.  These regions can also be
identified by contrasting panel~(a) with panel~(b), which shows the
photoionization rate density distribution in the L20N512pp simulation.
Sources in this simulation are located on the face of the box; as a
result, they are not correlated with density, and cells with enhanced
photoionization rate then move closer to the mean density.  Panels (c)
and (d) show the effect of removing cells containing sources from the
fiducial simulation, for different values of $a$ and $b$.  As
expected, cells with high photoionization rate vanish when we exclude
the sources and their neighbouring cells.  The curves in panels~(b),
(c), and (d) show the \emph{median} photoionization rates as a
function of density.  The red curves in panels (c) and (d) show the
median curve from panel (b) for easy comparison, while black curves in
panels (c) and (d) show the medians from the respective
photoionization rate distributions.  At high densities, the
  photoionisation rate distribution in panel~(a) is bimodal; we choose
  values of $a$ and $b$ that completely remove cells belonging to the
  high photoionization rate branch.  At the end of this procedure,
only the self-shielded cells, which lie in ionized regions, remain.
When this happens, the median photoionization rate agrees with that in
the corresponding plane-parallel simulation, as seen in panel~(d).  We
have chosen $a=0.86$ and $b=-4.14$ for the fiducial simulation at
$z=7$.  We identify the values of $a$ and $b$ for each of our
simulations at all redshifts separately.  In practice, the values are
not very different from those for the fiducial simulation.

\begin{figure*}
   \begin{center}
      \includegraphics[width=\textwidth]{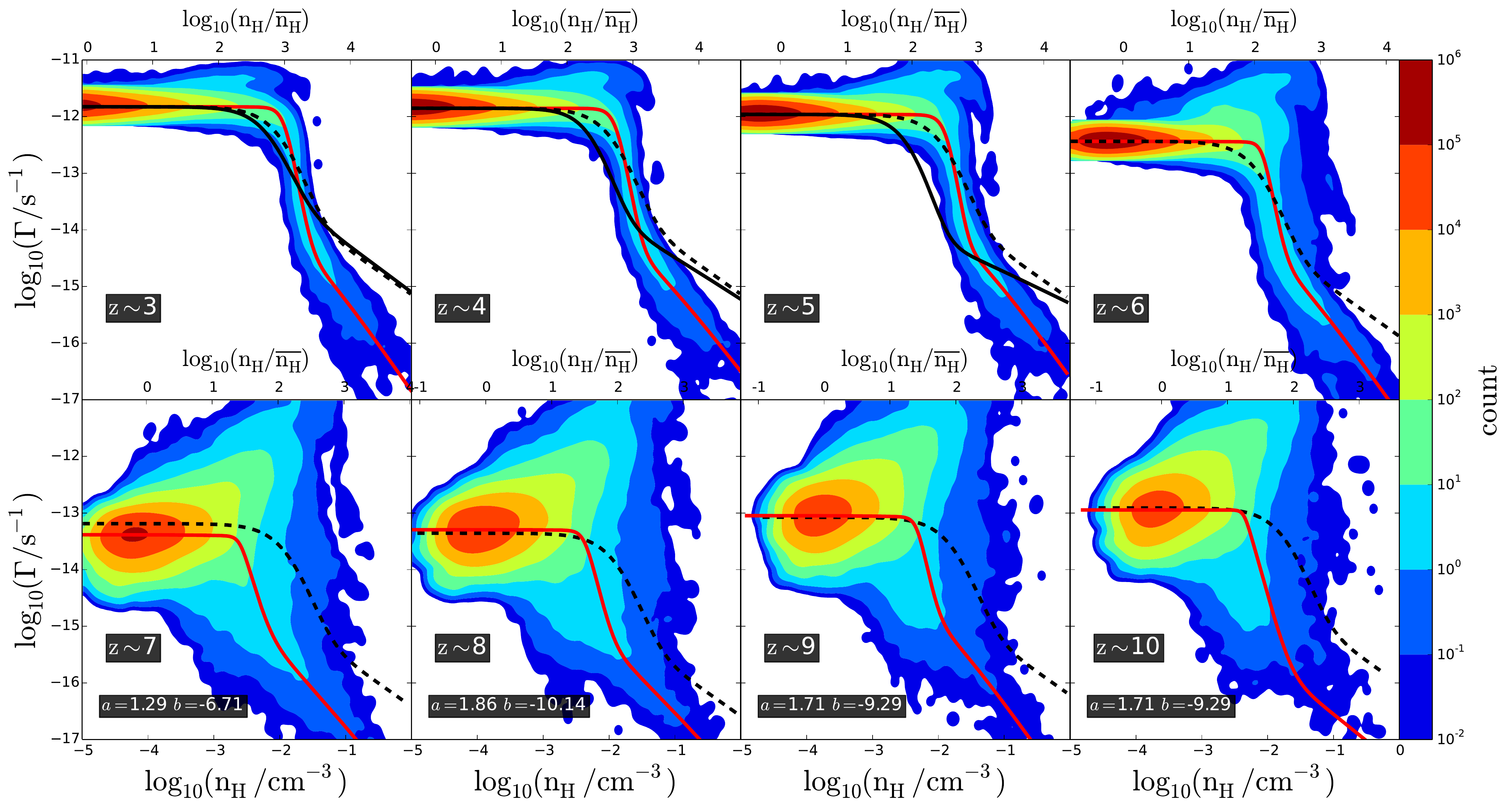}   
      \caption{The photoionization rate density distribution at
        various redshifts from the L10N512fs simulation.  In each
        panel, the red curve shows the best-fit curve of the form
        given by Equation~\ref{fit_Rahm}.  (Note that the curve is fit
        after masking regions near ionizing sources, as described in
        the text.  However the masked regions are still shown in
          the figure.)  The black solid curves show the fits
        presented by \citet{2013MNRAS.430.2427R} at $z<6$, while the
        black dashed curve shows their `average' fit.}
  \label{fig:fid}
  \end{center}
 \end{figure*}

\begin{figure*}
   \begin{center}
      \includegraphics[width=\textwidth]{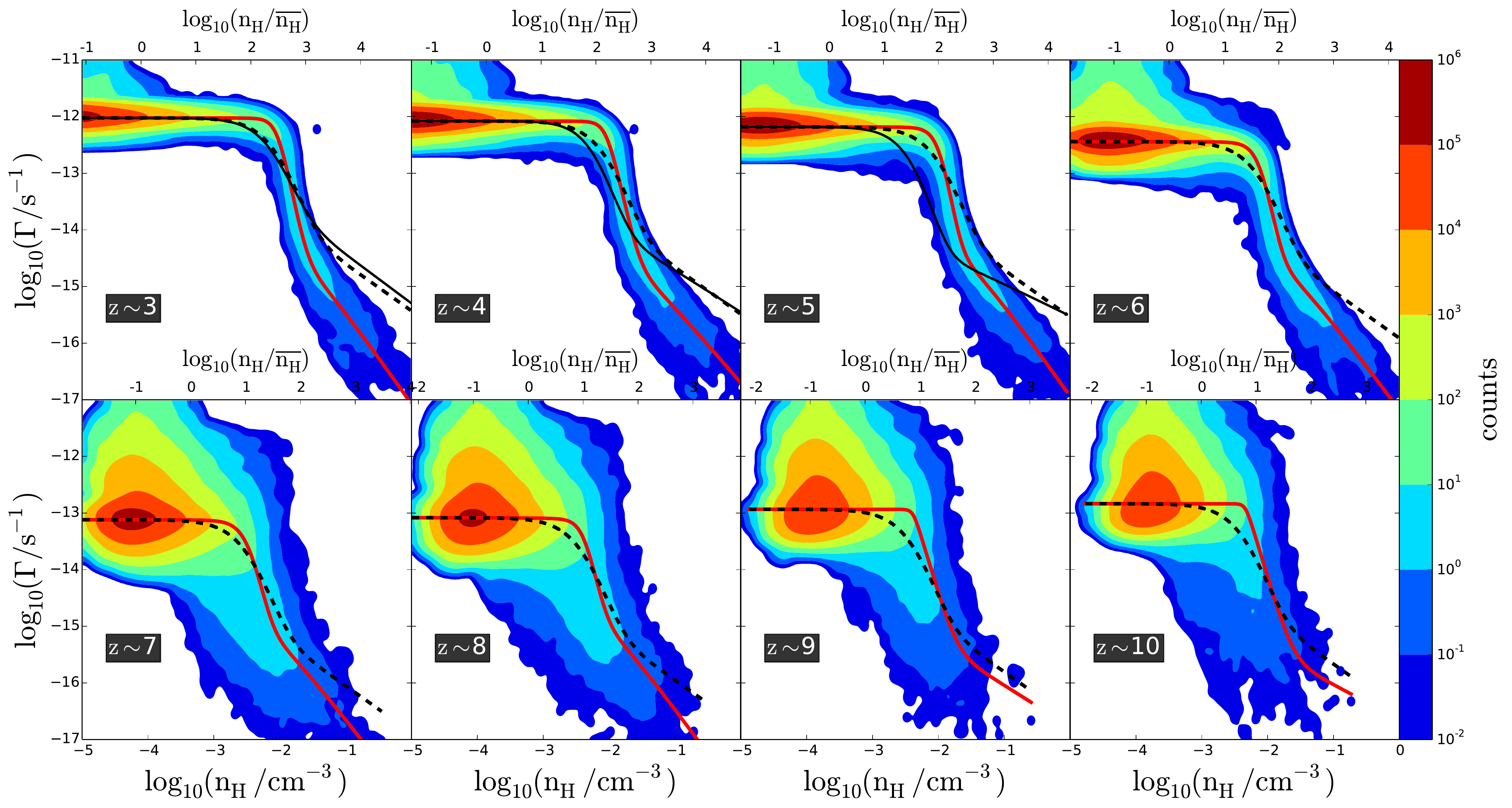}   
  \caption{The photoionization rate density distribution at various
    redshifts from the L10N512pp simulation.  The various curves shown
    are the same as in Figure~\ref{fig:fid}.}
    \label{fit:fid_pp}
  \end{center}
 \end{figure*}

\begin{figure*}
   \begin{center}
      \includegraphics[width=\textwidth,height=\textheight,keepaspectratio]{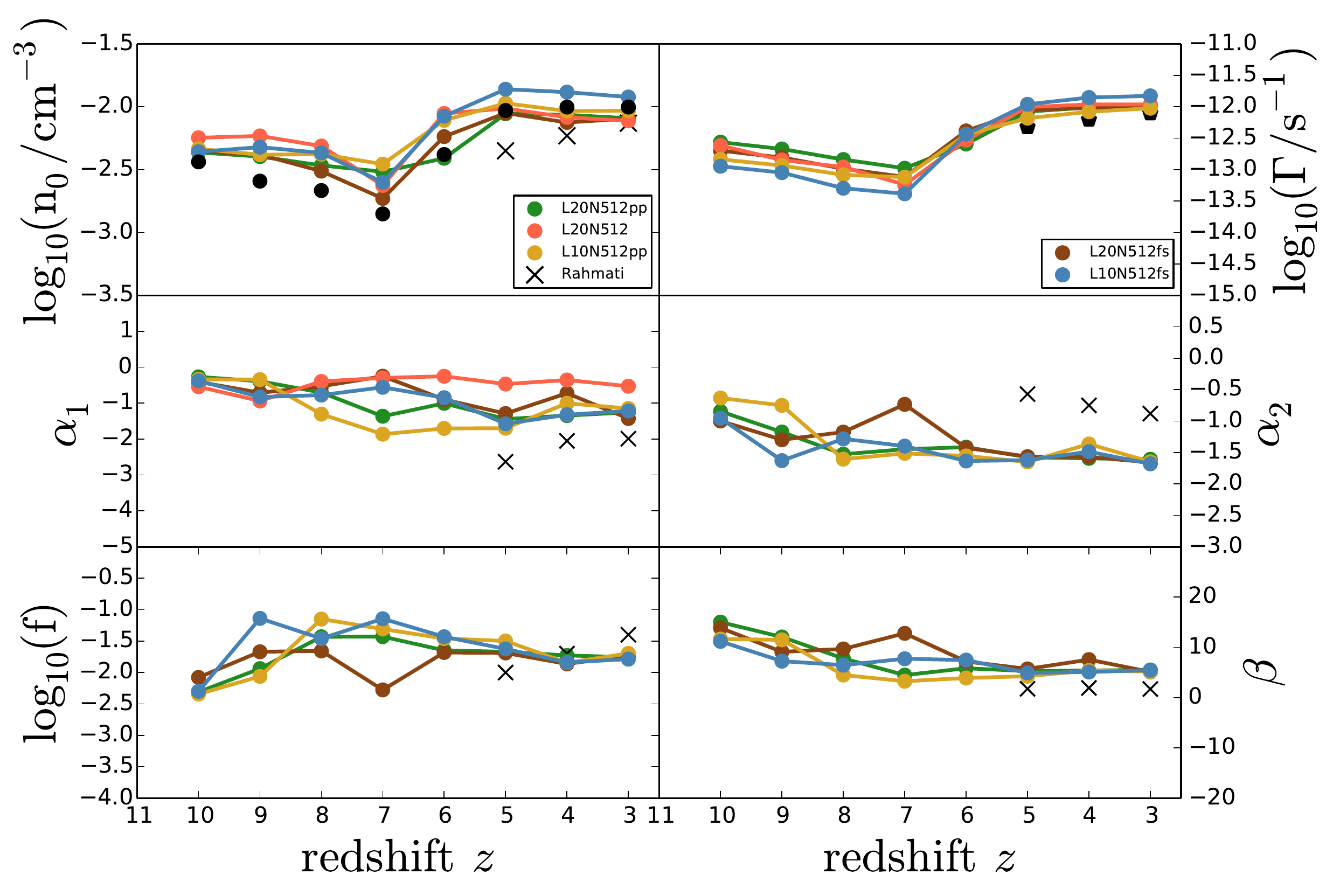}
  \caption{Evolution of the volume-averaged hydrogen photoionization
    rate in ionized regions, and of the five parameters of
    Equation~\ref{fit_Rahm} in all five simulations considered in this
    paper.  Black crosses show parameter values from
    \citet{2013MNRAS.430.2427R}.  In the photoionization rate panel,
    the black symbols show the photoionization rate inferred from the
    fits provided by \citet{2013MNRAS.430.2427R}, by using
    Equation~\ref{eqn:n0gamma}.  Similarly, black filled circles in
    the $n_0$ panel, show the self-shielding density given by
    Equation~\ref{eqn:n0gamma} for the fiducial simulation.  As
    discussed in the text, the L20N512 simulation does not constrain
    the slope parameters very well.  Therefore, this simulation is
    only shown in three of the six panels.}
    \label{fig:paramEvolAllSims}
  \end{center}
 \end{figure*}

Figure~\ref{fig:fid} shows the photoionization rate density
distributions from redshift $z=3$ to $z=10$ after the neutral regions
are removed following the procedure described above.  Note that cells
located near ionizing sources have not been masked yet.  In
Figure~\ref{fig:fid}, we show results from the L10N512fs simulation
instead of our fiducial L20N512 simulation, as the former simulation
better illustrates the behaviour of the photoionization rate at high
densities, due to its higher resolution.  As we will discuss below,
the results are nearly identical in the two simulations.  The average
photoionization rate follows the evolution seen in
Figure~\ref{xion_HI_vs_z}: there is little evolution from $z=10$ to
$7$ and then a jump by an order of magnitude from $\Gamma\sim
10^{-13}~\mathrm{s}^{-1}$ to $\Gamma\sim 10^{-12}~\mathrm{s}^{-1}$ at
$z=6$.  The density dependence of the photoionization rate has a
characteristic shape that remains essentially unchanged throughout
reionization: low density regions have high photoionization rates and
high density regions (that are not close to ionizing sources) have low
photoionization rates.  This shape is caused by the shielding.  As a
uniform photoionization rate background gets established after the
completion of reionization, the photoionization rate distribution in
Figure~\ref{fig:fid} narrows.

The photoionization rate distribution shown in Figure~\ref{fig:fid}
can now be used to understand self-shielding in our simulations.  We
find that, at all redshifts studied here, the photoionization rate has
a background value at low densities that does not depend on the gas
density.  The spread around the background value is quite small at low
redshifts but is larger at high redshifts, before individual ionized
regions have percolated.  At high densities ($n_\mathrm{H}\gtrsim
5\times 10^{-3}$~cm$^{-3}$), the photoionization rate drops below its
background value due to self-shielding.  At even higher densities
($n_\mathrm{H}\gtrsim 5\times 10^{-2}$~cm$^{-3}$), the photoionization
rate increases further, due to recombination radiation.  This causes
the photoionization rate density curve to flatten at very high
densities.

We find that the photoionization rate distribution is described very
well by the fitting function introduced by
\citet{2013MNRAS.430.2427R},
\begin{equation}
  \frac{\Gamma}{\langle\Gamma\rangle_{\mathrm{HII}}}=
  (1-f)\left[1+\left(\frac{n_H}{n_0}\right)^\beta\right]
  ^{\alpha_1}+f\left[1+\frac{n_H}{n_0}\right]^{\alpha_2}.
\label{fit_Rahm}
\end{equation}
In this fitting function, the parameter $n_0$ describes the
characteristic density at which self-shielding occurs.  The parameters
$\beta$ and $\alpha_1$ describe the slope of the photoionization rate
density curve caused by self-shielding, while the parameters $f$ and
$\alpha_2$ describe the enhanced slope of the photoionization rate
density curve at the highest densities.  We fit this form to the
photoionization rate distributions of Figure~\ref{fig:fid}, after
masking the sources following the procedure described above.  This fit
is illustrated by the solid red curves in Figure~\ref{fig:fid}.
Figure~\ref{fit:fid_pp} shows the corresponding fits in the
plane-parallel source model (L10N512pp).  The black solid curves shows
the result from comparable simulations by \citet{2013MNRAS.430.2427R}
at redshifts $z<6$.  \citet{2013MNRAS.430.2427R} noted that the
parameters in Equation~(\ref{fit_Rahm}) do not seem to evolve
significantly between redshifts $z=1$ and $6$ in their simulations.
Therefore, they put forward a single set of parameter values to
describe the photoionization rate density curve in this redshift
range.  The black dashed curve in all panels of Figures~\ref{fig:fid}
and \ref{fit:fid_pp} shows this ``average'' curve.

The photoionization rate distributions in Figures~\ref{fig:fid} and
\ref{fit:fid_pp} have several interesting features.  In both cases, at
all redshifts, Equation~(\ref{fit_Rahm}) describes the distributions
very well.  In the L10N512fs model (Figure~\ref{fig:fid}), in which
ionizing sources are placed at the locations of haloes, there is an
enhancement in the photoionization rate at high densities due to the
presence of sources in this regions, as discussed above.  In the
plane-parallel L10N512pp simulation, this enhancement is no longer
seen; instead, at lower densities, there is a spread in the
photoionization rate around the background values.  This spread is due
to the enhancement of the photoionization rate in cells that are close
to the sides of the simulation box, where the ``sources'' of the
plane-parallel fronts are located.  As most of the cells have low
densities (close to the mean) the spread in the photoionization rate
also moves to these densities.  The difference between the fiducial
and the plane-parallel models is also reflected at the high density
end of this plot.  We find that the high density part is better
sampled in the plane-parallel simulation than in the fiducial
simulation because in the latter, many of these cells are transferred
to the high photoionization rate part of the plot due to the presence
of ionizing sources in them.  This is reflected in the relatively poor
fits at high densities in the L10N512fs simulation, compared to the
L10N512pp simulation.  Note that in all of our simulations, the high
density cells are also affected by star formation.  Finally, our
simulations corroborate the findings of \citet{2013MNRAS.430.2427R} by
also showing very little evolution in self-shielding across redshifts,
even at redshifts before reionization.  Nonetheless, there are
differences between our best-fit model and that of
\citet{2013MNRAS.430.2427R}: the characteristic slope of the
self-shielding transition is much steeper in both our restricted-halo
and plane-parallel simulations.  As we discuss in greater detail
below, this difference arises due to a combination of three reasons:
(a) difference in the radiative transfer method, (b) star formation,
and (c) recombination radiation.  \citet{2013MNRAS.430.2427R}
highlight the role of different radiative transfer algorithms by
noting a similar difference between their results and those from
ray-tracing radiative transfer simulations by
\citet{2011ApJ...737L..37A}.  We have already mentioned that the star
formation prescription used in our simulations may deplete regions
with high gas density, thereby affecting the results shown in
Figures~\ref{fig:fid} and \ref{fit:fid_pp}.  Finally, the effect of
recombinations is likely underestimated in our simulations due to
their monochromatic nature, as discussed in Section~\ref{postRT}.
The temperature evolution in our simulations is unlikely to have
  an effect on the photoionisation rate distribution due to the small
  pressure smoothing scale at these redshifts
  \citep{2015ApJ...812...30K}.

The redshift evolution of the parameters in Equation~(\ref{fit_Rahm})
is summarised in Figure~\ref{fig:paramEvolAllSims} for all five of our
simulations.  Also shown in the top right panel of this figure is the
evolution of the volume-averaged photoionization rate in ionized
regions.  Red points in Figure~\ref{fig:paramEvolAllSims} show results
from our fiducial simulations for redshifts $z=3$--$10$.  All
simulations agree with each other to a reasonable degree, particularly
in the evolution of the characteristic self-shielding density $n_0$.
The black crosses in Figure~\ref{fig:paramEvolAllSims} show the
corresponding values from \citet{2013MNRAS.430.2427R} for $z<6$.  We
find that the self-shielding density $n_0$ stays roughly constant at
$n_0\sim 10^{-2}$~cm$^{-3}$ at post-reionization redshifts ($z<6$).
This density corresponds to overdensities of about 100.  At higher
redshifts, the self-shielding density is lower, $n_0\sim
10^{-3}$~cm$^{-3}$, which corresponds to overdensities of about 10.
It is apparent from comparing the top two panels of
Figure~\ref{fig:paramEvolAllSims} that the characteristic
self-shielding density $n_0$ follows the evolution in the
photoionization rate at all redshifts.  As discussed above, and
evidenced by Figures~\ref{fig:fid} and \ref{fit:fid_pp}, the
self-shielding density in our simulations is close to that in the
simulations of \citet{2013MNRAS.430.2427R}.  The self-shielding
density can be written as a function of the photoionization rate as
\citep{2001ApJ...559..507S, 2005ApJ...622....7F, 2013MNRAS.430.2427R},
\begin{multline}
  n_0\sim 6.73\times 10^{-3}\mathrm{cm}^{-3}\left(\frac{\sigma_{\nu_\mathrm{HI}}}{2.49\times 10^{-18}\mathrm{cm}^2}\right)^{-2/3}\\
  \times T_4^{0.17}\Gamma_{-12}^{2/3}\left(\frac{f_g}{0.17}\right)^{-1/3},
  \label{eqn:n0gamma}
\end{multline}
where $T_4=T/10^4$, $\Gamma_{-12}=\Gamma/10^{-12}$~s$^{-1}$, and
$f_g=\Omega_b/\Omega_m$ is the cosmic baryon fraction.  This relation
assumes absorbers with temperature $T\sim 10^4$~K, column density
$N_\mathrm{HI}\sim 1/\sigma$, and a typical size given by the Jeans
scale.  The black points in the top left panel of
Figure~\ref{fig:paramEvolAllSims} show the $n_0$ derived from our
volume-averaged photoionization rate from ionized regions using
Equation~(\ref{eqn:n0gamma}) for the fiducial simulation.  We find
that Equation~(\ref{eqn:n0gamma}) describes the evolution in $n_0$
reasonably well.  At low redshifts ($z<6$) the agreement is very good.
At high redshifts there is a constant offset due to the difference in
the cell selection for the photoionization rate and self-shielding
density.  The average values of parameters shown in
Figure~\ref{fig:paramEvolAllSims} are presented in
Appendix~\ref{sec:param_values} as a function of redshift.  These
should be useful for use in simulations without radiative transfer
that rely on approximate models for self-shielding.

The slopes of the photoionization rate density distribution also agree
quite well between various simulations.  The fiducial simulation
(L20N512) is unable to resolve the highest density part of the
distribution very well, partly due to the low spatial resolution and
partly because the sources in this simulation occupy a wide dynamic
range in halo mass.  Therefore, Figure~\ref{fig:paramEvolAllSims} only
shows one of the slopes, $\alpha_1$, from this simulation.  There are
also differences between the slopes in our simulations and those of
\citet{2013MNRAS.430.2427R} at post-reionization redshifts.  As
discussed by \citet{2013MNRAS.430.2427R}, this is partly due to the
difference in the radiative transfer methods, but also due to the
difference in the way recombination radiation is treated.

\begin{figure*}
  \begin{center}
    \includegraphics[width=\textwidth,height=\textheight,keepaspectratio]{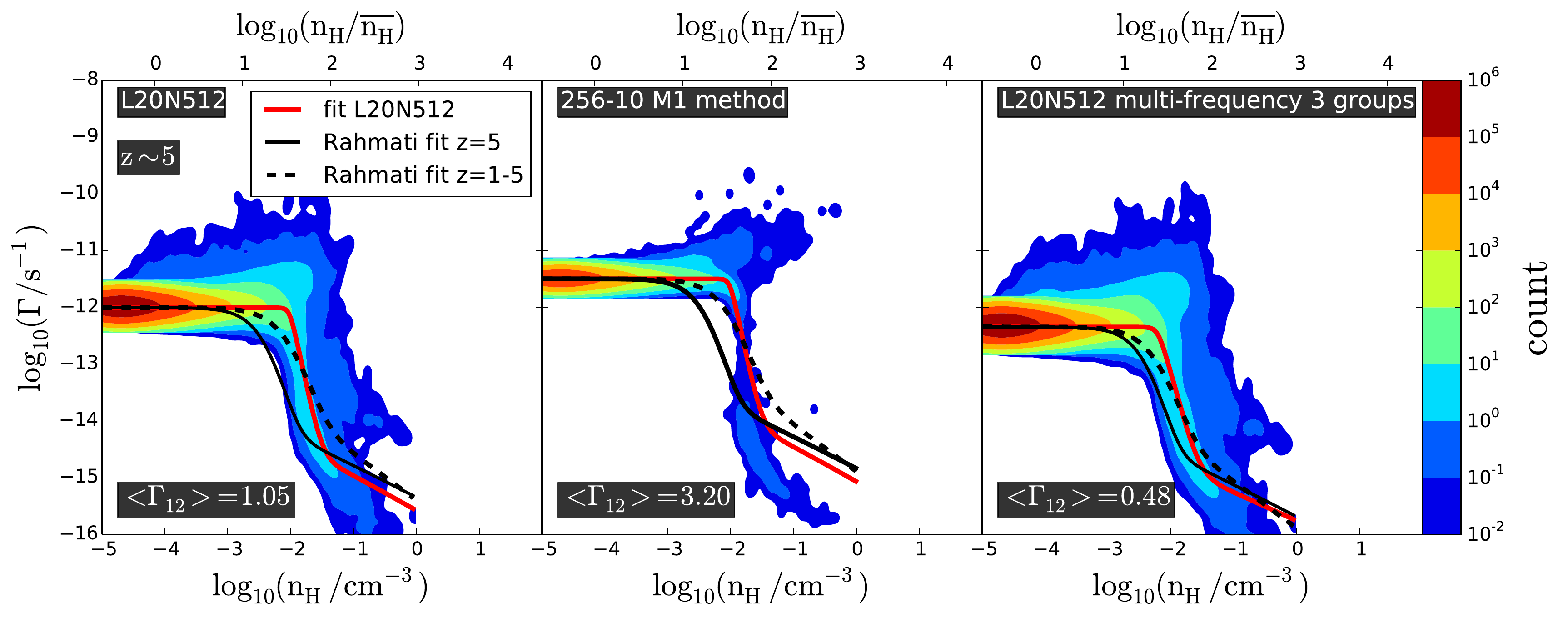}   
    \caption{Comparison of our results with those from two other
      simulations.  The left panel shows the photoionization rate
      distribution at $z=5$ from our L20N512 simulation.  The middle
      panel shows the photoionization rate distribution at the same
      redshift when the radiative transfer calculation is performed
      with the same density and source fields but using the
      moment-based implementation by \citet{2015MNRAS.453.3593B}.
      The right panel shows results from a multi-frequency
        simulation using three frequency bins.  In all three panels,
      the red curve shows the best-fit curve of the form given by
      Equation~\ref{fit_Rahm} from left panel.  The solid and dashed
      black curves in all three panels show two photoionization rate
      distribution fits from \citet{2013MNRAS.430.2427R}.}
    \label{comparison}
  \end{center}
\end{figure*}

\section{Discussion}
\label{discussion}

The self-shielding of intergalactic hydrogen described in the previous
section is potentially sensitive to several assumptions made in our
simulations.  We now discuss some of the more important of these.

Figure~\ref{fig:paramEvolAllSims} highlights the agreement between all
of our simulations, quantified using the fitting function in
Equation~(\ref{fit_Rahm}).  As discussed above, as in the fiducial
simulation, the self-shielding density $n_0$ stays roughly constant at
$10^{-2}$~cm$^{-3}$ at post-reionization redshifts ($z<6$) and tracks
the photoionization rate evolution at higher redshifts.  The
self-shielding density agrees between the simulations to within a
factor of 2.  This level of discrepancy is completely explained by the
differences in the photoionization rate in these simulations, as seen
in Figure~\ref{xion_HI_vs_z}.  A similar level of agreement is seen in
the parameters $\beta$ and $\alpha_1$, which describe the
characteristic slope of the photoionization rate density relation
caused by self-shielding.  There is significant improvement in the
parameters, $f$ and $\alpha_2$, which predominantly describe the
modification of the slope of the photoionization rate density relation
at high densities due to recombination radiation.  This is because the
high densities are much better resolved in the higher resolution
simulations than in the fiducial simulation
\citep{2013MNRAS.430.2427R}.

It has been noted in the literature that self-shielding in simulations
can also be affected by the radiative transfer method used.  In this
paper we have used a moment-based algorithm of radiative transfer, as
implemented in \textsc{aton} (see Section~2.1).  We have performed
monochromatic radiative transfer but self-consistently include the
recombination radiation.  We do not use the on-the-spot approximation.
In contrast, for example, \citet{2013MNRAS.430.2427R} used the
\textsc{traphic} code \citep{2008MNRAS.389..651P} in their work, where
radiative transfer is implemented by tracing photon packets.  This
technique is very different from the moment-based method used in the
present paper.  \citet{2011ApJ...737L..37A} used a ray-tracing
radiative transfer algorithm in the \textsc{owls} simulations.  These
authors ignore recombination radiation.  \citet{2013MNRAS.430.2427R}
note that self-shielding in \textsc{owls} is somewhat different from
their simulation at least at post-reionization redshifts, although
overall the agreement is good.  The characteristic slope of the
photoionization rate density relation is somewhat steeper in the ray
tracing case as compared to the photon packet method.  This could
partly be a resolution effect and partly be due to inaccuracies in
calculating the absorption of UV photons at large scales in the
ray-tracing codes.  In Figure~\ref{comparison}, we compare our result
from the fiducial 20~cMpc$/h$ box with a similar simulation that uses
the moment-based radiative transfer as implemented by
\citet{2015MNRAS.453.3593B}.  This simulation is performed on a
  $256^3$ grid in a 10~cMpc$/h$ box so that it has the same spatial
  resolution as our fiducial simulation.  The radiative transfer
  algorithm used by \citet{2015MNRAS.453.3593B} uses the M1 closure
  similar to our radiative transfer method. Both of these simulations
are at redshift $z=5$.  They use similar density and source fields,
and are calibrated in similar way.  The solid red curve in
Figure~\ref{comparison} shows the fit to the photoionization rate
density distribution from the fiducial L20N512 simulation.  But the
middle panel of Figure~\ref{comparison} shows that this provides a
good description also of the results from a different radiative
transfer implementation.  It is also interesting that both simulations
have steeper characteristic slopes as compared to the results of
\citet{2013MNRAS.430.2427R}: the dashed black curve shows the average
fit from their work while the solid black curve shows their fit at
redshift $z=5$. Both curves have shallower characteristic slopes that
in our simulations at this redshift.  This suggests that the shallower
slopes of \citet{2013MNRAS.430.2427R} could be a result of limited
resolution.

\begin{figure}
   \begin{center}
      \includegraphics[width=\columnwidth]{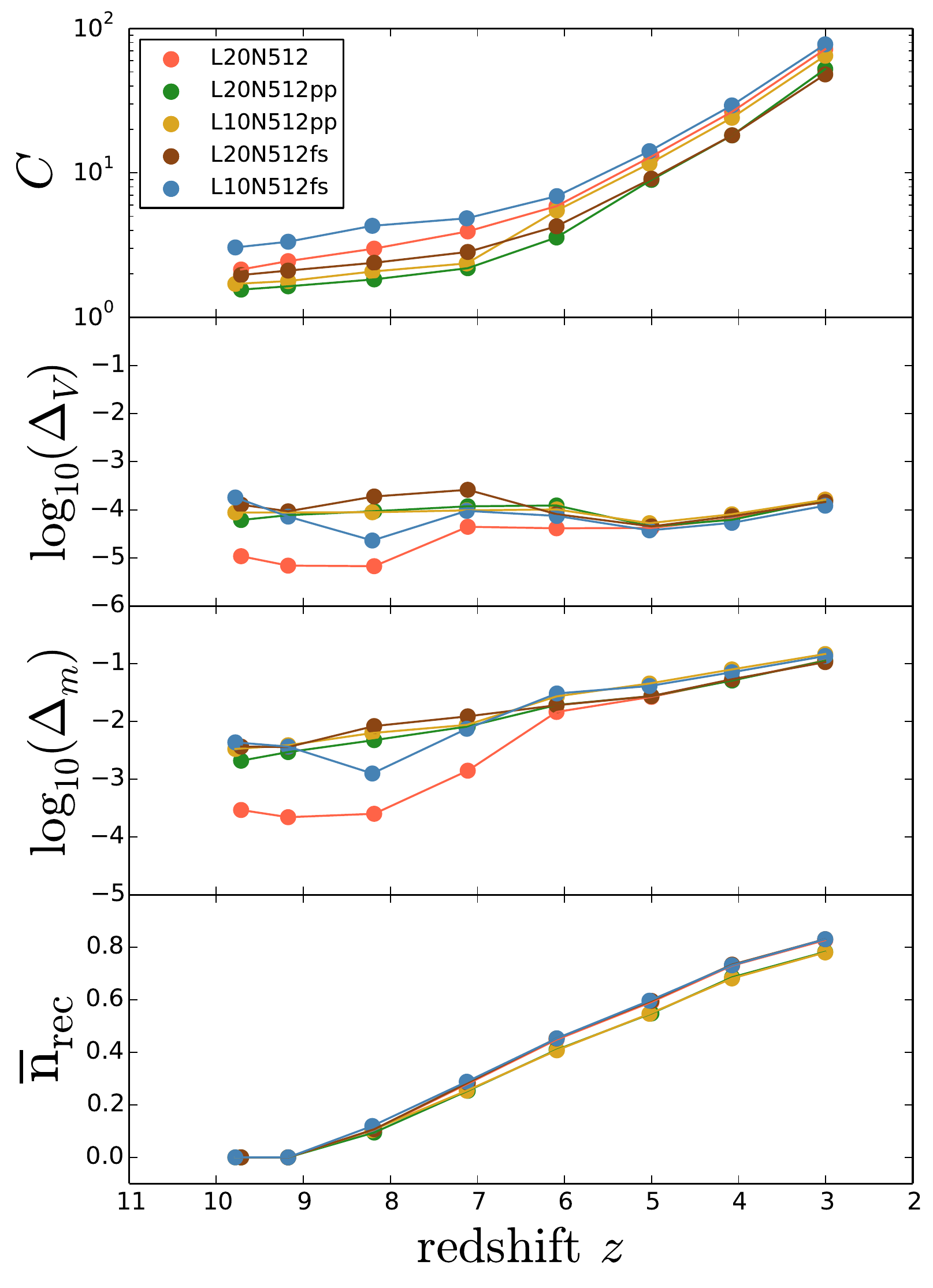}
  \caption{Evolution of the clumping factor, the volume and mass
    fraction of self-shielded regions, and the mean number of
    recombinations per hydrogen atom in the five simulations presented
    in this paper.}
    \label{fig:clump}
  \end{center}
\end{figure}

The third panel of Figure~\ref{comparison} explores the effect of
monochromaticity in our simulations.  It shows the photoionization
rate density distribution at redshift $z=5$ from a multi-frequency
counterpart of our fiducial simulation.  This simulation uses three
photon energy bins at 17.87~eV, 30.23~eV, and 59.38~eV, but otherwise
has identical properties as our fiducial simulation.
Figure~\ref{comparison} shows that the resultant picture of
self-shielding emerging from this multi-frequency simulation is
remarkably similar to that in our monochromatic simulation.  The red
curve in this panel shows our best-fit model from our L20N512
simulation at $z=5$.  The self-shielding density and the
characteristic slope are in agreement with the multi-frequency
simulation.  (The normalization of the curve has been adjusted to
  match with the somewhat different value of the average
  photoionization rate, $\langle\Gamma\rangle_\mathrm{HII}$, which has
  values of $10^{-12}$~s$^{-1}$, $3\times 10^{-12}$~s$^{-1}$, and $6
  \times 10^{-13}$~s$^{-1}$, in the three simulations shown in
  Figure~\ref{comparison}, respectively.)  The red curve, which
describes the simulation in the left panel of Figure~\ref{comparison}
also provides a good description of the photoionisation rate
distribution in the central and right panels of this figure.  The best
fit parameters for the three simulations agree to within 10\%.

Finally, in Figure~\ref{fig:clump}, we show the evolution of the
clumping factor, the mass and volume fractions contained in the
self-shielded regions, and the average number of recombinations per
hydrogen atom in our simulations.  The clumping factor is defined here
as $C=\langle n_\mathrm{HII}^2\rangle/\langle n_\mathrm{HII}\rangle^2$
and is evaluated only in ionized regions.  As expected the clumping
factor grows with time as nonlinear structure develops.  During the
epoch of reionization ($z\sim 6$--$10$), the clumping factor is low,
about 2--3.  This is in agreement with previous results from
\citet{2009MNRAS.394.1812P} and \citet{2006MNRAS.369.1625I}.  Further,
as the middle panel of Figure~\ref{fig:clump} shows, the self-shielded
systems occupy a very small ($<0.1\%$) fraction of the volume.  The
fractional mass content is relatively larger, about 1\% at $z\sim 6$,
decreasing to $0.1\%$ up to $z\sim 10$.  The decrease in the mass
fraction towards high redshifts is due to the increase in the
photoionization rate in high density regions.  Note that the fiducial
simulation, L20N512, shows considerably different behaviour in the
evolution of the mass and volume fractions and the mean number of
recombinations per hydrogen atom, due to the poor resolution of this
simulation in this regime, as discussed above.

A better measure of the effect of clumping on how reionization
proceeds, is the average number of recombinations per hydrogen atom in
ionized regions, defined as
\begin{equation}
  \bar n_\mathrm{rec}(t) = \int_0^t dt\alpha_\mathrm{B}(t)\left\langle\frac{n_\mathrm{HII}(t)^2}{n_\mathrm{H}(t)}\right\rangle
  \label{eqn:nrec}
\end{equation}
where $\alpha_\mathrm{B}$ is the case-B recombination rate of ionized
hydrogen, which evolves due to the evolution of gas temperature, and
the average is over ionized regions.  There is relatively little
difference between the mean number of recombinations in various
simulations, as most differences are at high redshifts where the
contribution to the integral in Equation~(\ref{eqn:nrec}) is small.
The value of $\bar n_\mathrm{rec}$ is somewhat smaller than previous
studies \citet{2006MNRAS.369.1625I}.  This is partly due to enhanced
self-shielding in our simulations due to relatively high resolution.
This reduces the number of high density ionized regions, thereby
reducing the number of recombinations.

\section{Conclusions}
\label{prospects}

We have used here full radiative transfer simulations of the epoch of
hydrogen reionization carefully calibrated with \lya forest data to
study the self-shielding of hydrogen during and after reionization. We
find that in spite of large fluctuations in the photoionization rate
during reionization, the characteristic density above which
self-shielding is important follows the evolution of the average
photoionization rate in a manner similar to that in the
post-reionization Universe. This results in similar typical values for
the characteristic self-shielding density, of $10^2$--$10^3$ cm$^{-3}$
for a wide range of redshifts from z = 10 to z = 2. This corresponds
to overdensities of $\sim 100$ in the post-reionization Universe, but
more like $\sim 10$ during reionization, reflecting the fact that
during reionization photoionization rates are typically low enough
that the filaments in the cosmic web are often self-shielded.

We have shown our findings to be robust against various assumptions
made in our study and have shown that our inferred values of the
self-shielding density have converged with respect to numerical
resolution in our simulation.  The detailed properties of the ionizing
sources are not critically important as long as they result in a
similar evolution of the photoionization rate and ionized gas
fraction. The results from our mono-frequency radiative transfer
simulations are in good agreement with those from equivalent
multi-frequency radiative transfer simulations.  Our results are
  also in good overall agreement with previous studies of the
  post-reionization IGM \citep{2013MNRAS.430.2427R}, but with a
  steeper slope of the self-shielding transition.

To obtain our results we have developed a method to efficiently
isolate the ionized regions in our simulations, which should have
wider applications in the study of reionization simulations and for
the analysis of future datasets that will yield a three-dimensional
tomographic view of the epoch of reionization by measuring the
brightness of the 21 cm line from the IGM.

Although self-shielded regions occupy a very small fraction of the
cosmological volume, they nevertheless contain about 0.1--1\% of the
total hydrogen mass during the epoch of reionization in our
simulation.  This has an effect on the recombination rate of hydrogen,
which can be quantified by the average number of recombinations per
hydrogen atom.  The typical value of this quantity in our simulations
during reionization is less than 0.4, slightly smaller that previous
studies.  Self-shielded regions of neutral hydrogen are known to
critically affect the visibility of \lya emitters during the epoch of
reionization as well as the clustering properties of the 21 cm signal
on large scales.  Modelling self-shielding accurately is thus
important.  We have provided parameter values at high redshifts
  for the fitting formula proposed by \citet{2013MNRAS.430.2427R} for
the self-shielding transition for use in simulations without radiative
transfer that rely on approximate models of self-shielding.

\section*{Acknowledgments}

We thank James Bolton, Tirthankar Roy Choudhury, Harley Katz, Laura
Keating, and Ewald Puchwein for helpful discussions.  This work was
supported by the ERC Advanced Grant 320596 ``The Emergence of
Structure during the epoch of Reionization".  The RAMSES simulations
presented in this paper were performed on the COSMOS Shared Memory
system at DAMTP, University of Cambridge operated on behalf of the
STFC DiRAC HPC Facility.  This equipment is funded by BIS National
E-infrastructure capital grant ST/J005673/1 and STFC grants
ST/H008586/1, ST/K00333X/1.  The ATON radiative transfer simulations
in this work were performed using the Wilkes GPU cluster at the
University of Cambridge High Performance Computing Service
(http://www.hpc.cam.ac.uk/), provided by Dell Inc., NVIDIA and
Mellanox, and part funded by STFC with industrial sponsorship from
Rolls Royce and Mitsubishi Heavy Industries.

\bibliographystyle{mnras}
\bibliography{biblio}

\begin{thebibliography}{}
\makeatletter
\relax
\def\mn@urlcharsother{\let\do\@makeother \do\$\do\&\do\#\do\^\do\_\do\%\do\~}
\def\mn@doi{\begingroup\mn@urlcharsother \@ifnextchar [ {\mn@doi@}
  {\mn@doi@[]}}
\def\mn@doi@[#1]#2{\def\@tempa{#1}\ifx\@tempa\@empty \href
  {http://dx.doi.org/#2} {doi:#2}\else \href {http://dx.doi.org/#2} {#1}\fi
  \endgroup}
\def\mn@eprint#1#2{\mn@eprint@#1:#2::\@nil}
\def\mn@eprint@arXiv#1{\href {http://arxiv.org/abs/#1} {{\tt arXiv:#1}}}
\def\mn@eprint@dblp#1{\href {http://dblp.uni-trier.de/rec/bibtex/#1.xml}
  {dblp:#1}}
\def\mn@eprint@#1:#2:#3:#4\@nil{\def\@tempa {#1}\def\@tempb {#2}\def\@tempc
  {#3}\ifx \@tempc \@empty \let \@tempc \@tempb \let \@tempb \@tempa \fi \ifx
  \@tempb \@empty \def\@tempb {arXiv}\fi \@ifundefined
  {mn@eprint@\@tempb}{\@tempb:\@tempc}{\expandafter \expandafter \csname
  mn@eprint@\@tempb\endcsname \expandafter{\@tempc}}}

\bibitem[\protect\citeauthoryear{{Altay}, {Theuns}, {Schaye}, {Crighton}  \&
  {Dalla Vecchia}}{{Altay} et~al.}{2011}]{2011ApJ...737L..37A}
{Altay} G.,  {Theuns} T.,  {Schaye} J.,  {Crighton} N.~H.~M.,   {Dalla Vecchia}
  C.,  2011, \mn@doi [\apjl] {10.1088/2041-8205/737/2/L37}, \href
  {http://adsabs.harvard.edu/abs/2011ApJ...737L..37A} {737, L37}

\bibitem[\protect\citeauthoryear{{Altay}, {Theuns}, {Schaye}, {Booth}  \&
  {Dalla Vecchia}}{{Altay} et~al.}{2013}]{2013MNRAS.436.2689A}
{Altay} G.,  {Theuns} T.,  {Schaye} J.,  {Booth} C.~M.,   {Dalla Vecchia} C.,
  2013, \mn@doi [\mnras] {10.1093/mnras/stt1765}, \href
  {http://adsabs.harvard.edu/abs/2013MNRAS.436.2689A} {436, 2689}

\bibitem[\protect\citeauthoryear{{Anderson}, {Governato}, {Karcher}, {Quinn}
  \& {Wadsley}}{{Anderson} et~al.}{2016}]{2016arXiv160605352A}
{Anderson} L.,  {Governato} F.,  {Karcher} M.,  {Quinn} T.,   {Wadsley} J.,
  2016, preprint, \href {http://adsabs.harvard.edu/abs/2016arXiv160605352A} {}
  (\mn@eprint {arXiv} {1606.05352})

\bibitem[\protect\citeauthoryear{{Aubert} \& {Teyssier}}{{Aubert} \&
  {Teyssier}}{2008}]{2008MNRAS.387..295A}
{Aubert} D.,  {Teyssier} R.,  2008, \mn@doi [\mnras]
  {10.1111/j.1365-2966.2008.13223.x}, \href
  {http://cdsads.u-strasbg.fr/abs/2008MNRAS.387..295A} {387, 295}

\bibitem[\protect\citeauthoryear{{Baek}, {Di Matteo}, {Semelin}, {Combes}  \&
  {Revaz}}{{Baek} et~al.}{2009}]{2009A&A...495..389B}
{Baek} S.,  {Di Matteo} P.,  {Semelin} B.,  {Combes} F.,   {Revaz} Y.,  2009,
  \mn@doi [\aap] {10.1051/0004-6361:200810757}, \href
  {http://cdsads.u-strasbg.fr/abs/2009A\%26A...495..389B} {495, 389}

\bibitem[\protect\citeauthoryear{{Barkana} \& {Loeb}}{{Barkana} \&
  {Loeb}}{2001}]{2001PhR...349..125B}
{Barkana} R.,  {Loeb} A.,  2001, \mn@doi [\physrep]
  {10.1016/S0370-1573(01)00019-9}, \href
  {http://adsabs.harvard.edu/abs/2001PhR...349..125B} {349, 125}

\bibitem[\protect\citeauthoryear{{Barnett}, {Warren}, {Becker}, {Mortlock},
  {Hewett}, {McMahon}, {Simpson}  \& {Venemans}}{{Barnett}
  et~al.}{2017}]{2017arXiv170203687B}
{Barnett} R.,  {Warren} S.~J.,  {Becker} G.~D.,  {Mortlock} D.~J.,  {Hewett}
  P.~C.,  {McMahon} R.~G.,  {Simpson} C.,   {Venemans} B.~P.,  2017, preprint,
  \href {http://adsabs.harvard.edu/abs/2017arXiv170203687B} {} (\mn@eprint
  {arXiv} {1702.03687})

\bibitem[\protect\citeauthoryear{{Bauer}, {Springel}, {Vogelsberger}, {Genel},
  {Torrey}, {Sijacki}, {Nelson}  \& {Hernquist}}{{Bauer}
  et~al.}{2015}]{2015MNRAS.453.3593B}
{Bauer} A.,  {Springel} V.,  {Vogelsberger} M.,  {Genel} S.,  {Torrey} P.,
  {Sijacki} D.,  {Nelson} D.,   {Hernquist} L.,  2015, \mn@doi [\mnras]
  {10.1093/mnras/stv1893}, \href
  {http://adsabs.harvard.edu/abs/2015MNRAS.453.3593B} {453, 3593}

\bibitem[\protect\citeauthoryear{{Becker} \& {Bolton}}{{Becker} \&
  {Bolton}}{2013}]{2013MNRAS.436.1023B}
{Becker} G.~D.,  {Bolton} J.~S.,  2013, \mn@doi [\mnras]
  {10.1093/mnras/stt1610}, \href
  {http://adsabs.harvard.edu/abs/2013MNRAS.436.1023B} {436, 1023}

\bibitem[\protect\citeauthoryear{{Bolton} \& {Haehnelt}}{{Bolton} \&
  {Haehnelt}}{2007}]{2007MNRAS.382..325B}
{Bolton} J.~S.,  {Haehnelt} M.~G.,  2007, \mn@doi [\mnras]
  {10.1111/j.1365-2966.2007.12372.x}, \href
  {http://adsabs.harvard.edu/abs/2007MNRAS.382..325B} {382, 325}

\bibitem[\protect\citeauthoryear{{Bolton} \& {Haehnelt}}{{Bolton} \&
  {Haehnelt}}{2013}]{2013MNRAS.429.1695B}
{Bolton} J.~S.,  {Haehnelt} M.~G.,  2013, \mn@doi [\mnras]
  {10.1093/mnras/sts455}, \href
  {http://adsabs.harvard.edu/abs/2013MNRAS.429.1695B} {429, 1695}

\bibitem[\protect\citeauthoryear{{Calverley}, {Becker}, {Haehnelt}  \&
  {Bolton}}{{Calverley} et~al.}{2011}]{2011MNRAS.412.2543C}
{Calverley} A.~P.,  {Becker} G.~D.,  {Haehnelt} M.~G.,   {Bolton} J.~S.,  2011,
  \mn@doi [\mnras] {10.1111/j.1365-2966.2010.18072.x}, \href
  {http://adsabs.harvard.edu/abs/2011MNRAS.412.2543C} {412, 2543}

\bibitem[\protect\citeauthoryear{{Chardin}, {Aubert}  \& {Ocvirk}}{{Chardin}
  et~al.}{2012}]{2012A&A...548A...9C}
{Chardin} J.,  {Aubert} D.,   {Ocvirk} P.,  2012, \mn@doi [\aap]
  {10.1051/0004-6361/201219992}, \href
  {http://adsabs.harvard.edu/abs/2012A\%26A...548A...9C} {548, A9}

\bibitem[\protect\citeauthoryear{{Chardin}, {Aubert}  \& {Ocvirk}}{{Chardin}
  et~al.}{2014}]{2014A&A...568A..52C}
{Chardin} J.,  {Aubert} D.,   {Ocvirk} P.,  2014, \mn@doi [\aap]
  {10.1051/0004-6361/201322355}, \href
  {http://adsabs.harvard.edu/abs/2014A%26A...568A..52C} {568, A52}

\bibitem[\protect\citeauthoryear{{Chardin}, {Haehnelt}, {Aubert}  \&
  {Puchwein}}{{Chardin} et~al.}{2015}]{2015MNRAS.453.2943C}
{Chardin} J.,  {Haehnelt} M.~G.,  {Aubert} D.,   {Puchwein} E.,  2015, \mn@doi
  [\mnras] {10.1093/mnras/stv1786}, \href
  {http://adsabs.harvard.edu/abs/2015MNRAS.453.2943C} {453, 2943}

\bibitem[\protect\citeauthoryear{{Chardin}, {Puchwein}  \&
  {Haehnelt}}{{Chardin} et~al.}{2017}]{2017MNRAS.465.3429C}
{Chardin} J.,  {Puchwein} E.,   {Haehnelt} M.~G.,  2017, \mn@doi [\mnras]
  {10.1093/mnras/stw2943}, \href
  {http://adsabs.harvard.edu/abs/2017MNRAS.465.3429C} {465, 3429}

\bibitem[\protect\citeauthoryear{{Choudhury} \& {Ferrara}}{{Choudhury} \&
  {Ferrara}}{2005}]{2005MNRAS.361..577C}
{Choudhury} T.~R.,  {Ferrara} A.,  2005, \mn@doi [\mnras]
  {10.1111/j.1365-2966.2005.09196.x}, \href
  {http://adsabs.harvard.edu/abs/2005MNRAS.361..577C} {361, 577}

\bibitem[\protect\citeauthoryear{{Choudhury}, {Haehnelt}  \&
  {Regan}}{{Choudhury} et~al.}{2009}]{2009MNRAS.394..960C}
{Choudhury} T.~R.,  {Haehnelt} M.~G.,   {Regan} J.,  2009, \mn@doi [\mnras]
  {10.1111/j.1365-2966.2008.14383.x}, \href
  {http://cdsads.u-strasbg.fr/abs/2009MNRAS.394..960C} {394, 960}

\bibitem[\protect\citeauthoryear{{Choudhury}, {Puchwein}, {Haehnelt}  \&
  {Bolton}}{{Choudhury} et~al.}{2015}]{2015MNRAS.452..261C}
{Choudhury} T.~R.,  {Puchwein} E.,  {Haehnelt} M.~G.,   {Bolton} J.~S.,  2015,
  \mn@doi [\mnras] {10.1093/mnras/stv1250}, \href
  {http://ukads.nottingham.ac.uk/abs/2015MNRAS.452..261C} {452, 261}

\bibitem[\protect\citeauthoryear{{Eisenstein} \& {Hut}}{{Eisenstein} \&
  {Hut}}{1998}]{1998ApJ...498..137E}
{Eisenstein} D.~J.,  {Hut} P.,  1998, \mn@doi [\apj] {10.1086/305535}, \href
  {http://adsabs.harvard.edu/abs/1998ApJ...498..137E} {498, 137}

\bibitem[\protect\citeauthoryear{{Fan} et~al.,}{{Fan}
  et~al.}{2006}]{2006AJ....132..117F}
{Fan} X.,  et~al., 2006, \mn@doi [\aj] {10.1086/504836}, \href
  {http://cdsads.u-strasbg.fr/abs/2006AJ....132..117F} {132, 117}

\bibitem[\protect\citeauthoryear{{Furlanetto} \& {Oh}}{{Furlanetto} \&
  {Oh}}{2005}]{2005MNRAS.363.1031F}
{Furlanetto} S.~R.,  {Oh} S.~P.,  2005, \mn@doi [\mnras]
  {10.1111/j.1365-2966.2005.09505.x}, \href
  {http://adsabs.harvard.edu/abs/2005MNRAS.363.1031F} {363, 1031}

\bibitem[\protect\citeauthoryear{{Furlanetto}, {Schaye}, {Springel}  \&
  {Hernquist}}{{Furlanetto} et~al.}{2005}]{2005ApJ...622....7F}
{Furlanetto} S.~R.,  {Schaye} J.,  {Springel} V.,   {Hernquist} L.,  2005,
  \mn@doi [\apj] {10.1086/426808}, \href
  {http://adsabs.harvard.edu/abs/2005ApJ...622....7F} {622, 7}

\bibitem[\protect\citeauthoryear{{Gnedin}}{{Gnedin}}{2000}]{2000ApJ...535..530G}
{Gnedin} N.~Y.,  2000, \mn@doi [\apj] {10.1086/308876}, \href
  {http://adsabs.harvard.edu/abs/2000ApJ...535..530G} {535, 530}

\bibitem[\protect\citeauthoryear{{Haardt} \& {Madau}}{{Haardt} \&
  {Madau}}{2012}]{2012ApJ...746..125H}
{Haardt} F.,  {Madau} P.,  2012, \mn@doi [\apj] {10.1088/0004-637X/746/2/125},
  \href {http://adsabs.harvard.edu/abs/2012ApJ...746..125H} {746, 125}

\bibitem[\protect\citeauthoryear{{Iliev}, {Mellema}, {Pen}, {Merz}, {Shapiro}
  \& {Alvarez}}{{Iliev} et~al.}{2006}]{2006MNRAS.369.1625I}
{Iliev} I.~T.,  {Mellema} G.,  {Pen} U.,  {Merz} H.,  {Shapiro} P.~R.,
  {Alvarez} M.~A.,  2006, \mn@doi [\mnras] {10.1111/j.1365-2966.2006.10502.x},
  \href {http://cdsads.u-strasbg.fr/abs/2006MNRAS.369.1625I} {369, 1625}

\bibitem[\protect\citeauthoryear{{Kaurov} \& {Gnedin}}{{Kaurov} \&
  {Gnedin}}{2013}]{2013ApJ...771...35K}
{Kaurov} A.~A.,  {Gnedin} N.~Y.,  2013, \mn@doi [\apj]
  {10.1088/0004-637X/771/1/35}, \href
  {http://adsabs.harvard.edu/abs/2013ApJ...771...35K} {771, 35}

\bibitem[\protect\citeauthoryear{{Kaurov} \& {Gnedin}}{{Kaurov} \&
  {Gnedin}}{2014}]{2014ApJ...787..146K}
{Kaurov} A.~A.,  {Gnedin} N.~Y.,  2014, \mn@doi [\apj]
  {10.1088/0004-637X/787/2/146}, \href
  {http://adsabs.harvard.edu/abs/2014ApJ...787..146K} {787, 146}

\bibitem[\protect\citeauthoryear{{Keating}, {Haehnelt}, {Cantalupo}  \&
  {Puchwein}}{{Keating} et~al.}{2015}]{2015MNRAS.454..681K}
{Keating} L.~C.,  {Haehnelt} M.~G.,  {Cantalupo} S.,   {Puchwein} E.,  2015,
  \mn@doi [\mnras] {10.1093/mnras/stv2020}, \href
  {http://adsabs.harvard.edu/abs/2015MNRAS.454..681K} {454, 681}

\bibitem[\protect\citeauthoryear{{Kulkarni}, {Hennawi}, {O{\~n}orbe}, {Rorai}
  \& {Springel}}{{Kulkarni} et~al.}{2015}]{2015ApJ...812...30K}
{Kulkarni} G.,  {Hennawi} J.~F.,  {O{\~n}orbe} J.,  {Rorai} A.,   {Springel}
  V.,  2015, \mn@doi [\apj] {10.1088/0004-637X/812/1/30}, \href
  {http://adsabs.harvard.edu/abs/2015ApJ...812...30K} {812, 30}

\bibitem[\protect\citeauthoryear{{Kulkarni}, {Choudhury}, {Puchwein}  \&
  {Haehnelt}}{{Kulkarni} et~al.}{2016}]{2016MNRAS.463.2583K}
{Kulkarni} G.,  {Choudhury} T.~R.,  {Puchwein} E.,   {Haehnelt} M.~G.,  2016,
  \mn@doi [\mnras] {10.1093/mnras/stw2168}, \href
  {http://ukads.nottingham.ac.uk/abs/2016MNRAS.463.2583K} {463, 2583}

\bibitem[\protect\citeauthoryear{{McGreer}, {Mesinger}  \&
  {D'Odorico}}{{McGreer} et~al.}{2015}]{2015MNRAS.447..499M}
{McGreer} I.~D.,  {Mesinger} A.,   {D'Odorico} V.,  2015, \mn@doi [\mnras]
  {10.1093/mnras/stu2449}, \href
  {http://cdsads.u-strasbg.fr/abs/2015MNRAS.447..499M} {447, 499}

\bibitem[\protect\citeauthoryear{{McQuinn}, {Hernquist}, {Zaldarriaga}  \&
  {Dutta}}{{McQuinn} et~al.}{2007}]{2007MNRAS.381...75M}
{McQuinn} M.,  {Hernquist} L.,  {Zaldarriaga} M.,   {Dutta} S.,  2007, \mn@doi
  [\mnras] {10.1111/j.1365-2966.2007.12085.x}, \href
  {http://cdsads.u-strasbg.fr/abs/2007MNRAS.381...75M} {381, 75}

\bibitem[\protect\citeauthoryear{{McQuinn}, {Oh}  \&
  {Faucher-Gigu{\`e}re}}{{McQuinn} et~al.}{2011}]{2011ApJ...743...82M}
{McQuinn} M.,  {Oh} S.~P.,   {Faucher-Gigu{\`e}re} C.-A.,  2011, \mn@doi [\apj]
  {10.1088/0004-637X/743/1/82}, \href
  {http://adsabs.harvard.edu/abs/2011ApJ...743...82M} {743, 82}

\bibitem[\protect\citeauthoryear{{Mesinger}, {Aykutalp}, {Vanzella},
  {Pentericci}, {Ferrara}  \& {Dijkstra}}{{Mesinger}
  et~al.}{2015}]{2015MNRAS.446..566M}
{Mesinger} A.,  {Aykutalp} A.,  {Vanzella} E.,  {Pentericci} L.,  {Ferrara} A.,
    {Dijkstra} M.,  2015, \mn@doi [\mnras] {10.1093/mnras/stu2089}, \href
  {http://adsabs.harvard.edu/abs/2015MNRAS.446..566M} {446, 566}

\bibitem[\protect\citeauthoryear{{Miralda-Escud{\'e}}, {Haehnelt}  \&
  {Rees}}{{Miralda-Escud{\'e}} et~al.}{2000}]{2000ApJ...530....1M}
{Miralda-Escud{\'e}} J.,  {Haehnelt} M.,   {Rees} M.~J.,  2000, \mn@doi [\apj]
  {10.1086/308330}, \href {http://cdsads.u-strasbg.fr/abs/2000ApJ...530....1M}
  {530, 1}

\bibitem[\protect\citeauthoryear{{Mitra}, {Choudhury}  \& {Ferrara}}{{Mitra}
  et~al.}{2015}]{2015MNRAS.454L..76M}
{Mitra} S.,  {Choudhury} T.~R.,   {Ferrara} A.,  2015, \mn@doi [\mnras]
  {10.1093/mnrasl/slv134}, \href
  {http://adsabs.harvard.edu/abs/2015MNRAS.454L..76M} {454, L76}

\bibitem[\protect\citeauthoryear{{Mutch}, {Geil}, {Poole}, {Angel}, {Duffy},
  {Mesinger}  \& {Wyithe}}{{Mutch} et~al.}{2016}]{2016MNRAS.462..250M}
{Mutch} S.~J.,  {Geil} P.~M.,  {Poole} G.~B.,  {Angel} P.~W.,  {Duffy} A.~R.,
  {Mesinger} A.,   {Wyithe} J.~S.~B.,  2016, \mn@doi [\mnras]
  {10.1093/mnras/stw1506}, \href
  {http://adsabs.harvard.edu/abs/2016MNRAS.462..250M} {462, 250}

\bibitem[\protect\citeauthoryear{{Ota} et~al.,}{{Ota}
  et~al.}{2008}]{2008ApJ...677...12O}
{Ota} K.,  et~al., 2008, \mn@doi [\apj] {10.1086/529006}, \href
  {http://cdsads.u-strasbg.fr/abs/2008ApJ...677...12O} {677, 12}

\bibitem[\protect\citeauthoryear{{Pawlik} \& {Schaye}}{{Pawlik} \&
  {Schaye}}{2008}]{2008MNRAS.389..651P}
{Pawlik} A.~H.,  {Schaye} J.,  2008, \mn@doi [\mnras]
  {10.1111/j.1365-2966.2008.13601.x}, \href
  {http://adsabs.harvard.edu/abs/2008MNRAS.389..651P} {389, 651}

\bibitem[\protect\citeauthoryear{{Pawlik}, {Schaye}  \& {van
  Scherpenzeel}}{{Pawlik} et~al.}{2009}]{2009MNRAS.394.1812P}
{Pawlik} A.~H.,  {Schaye} J.,   {van Scherpenzeel} E.,  2009, \mn@doi [\mnras]
  {10.1111/j.1365-2966.2009.14486.x}, \href
  {http://adsabs.harvard.edu/abs/2009MNRAS.394.1812P} {394, 1812}

\bibitem[\protect\citeauthoryear{{Planck Collaboration}}{{Planck
  Collaboration}}{2016}]{2016A&A...596A.108P}
{Planck Collaboration} 2016, \mn@doi [\aap] {10.1051/0004-6361/201628897},
  \href {http://adsabs.harvard.edu/abs/2016A%26A...596A.108P} {596, A108}

\bibitem[\protect\citeauthoryear{{Planck Collaboration} et~al.,}{{Planck
  Collaboration} et~al.}{2014}]{2014A&A...571A..16P}
{Planck Collaboration} et~al., 2014, \mn@doi [\aap]
  {10.1051/0004-6361/201321591}, \href
  {http://cdsads.u-strasbg.fr/abs/2014A%26A...571A..16P} {571, A16}

\bibitem[\protect\citeauthoryear{{Puchwein}, {Bolton}, {Haehnelt}, {Madau},
  {Becker}  \& {Haardt}}{{Puchwein} et~al.}{2015}]{2015arXiv1410.1531P}
{Puchwein} E.,  {Bolton} J.~S.,  {Haehnelt} M.~G.,  {Madau} P.,  {Becker}
  G.~D.,   {Haardt} F.,  2015, MNRAS accepted, ArXiv e-prints 1410.1531, \href
  {http://adsabs.harvard.edu/abs/2014arXiv1410.1531P} {}

\bibitem[\protect\citeauthoryear{{Rahmati}, {Pawlik}, {Rai{\v c}evic}  \&
  {Schaye}}{{Rahmati} et~al.}{2013}]{2013MNRAS.430.2427R}
{Rahmati} A.,  {Pawlik} A.~H.,  {Rai{\v c}evic} M.,   {Schaye} J.,  2013,
  \mn@doi [\mnras] {10.1093/mnras/stt066}, \href
  {http://adsabs.harvard.edu/abs/2013MNRAS.430.2427R} {430, 2427}

\bibitem[\protect\citeauthoryear{{Robertson}, {Ellis}, {Furlanetto}  \&
  {Dunlop}}{{Robertson} et~al.}{2015}]{2015ApJ...802L..19R}
{Robertson} B.~E.,  {Ellis} R.~S.,  {Furlanetto} S.~R.,   {Dunlop} J.~S.,
  2015, \mn@doi [\apjl] {10.1088/2041-8205/802/2/L19}, \href
  {http://adsabs.harvard.edu/abs/2015ApJ...802L..19R} {802, L19}

\bibitem[\protect\citeauthoryear{{Rorai} et~al.,}{{Rorai}
  et~al.}{2017}]{2017arXiv170408366R}
{Rorai} A.,  et~al., 2017, preprint, \href
  {http://adsabs.harvard.edu/abs/2017arXiv170408366R} {} (\mn@eprint {arXiv}
  {1704.08366})

\bibitem[\protect\citeauthoryear{{Schaye}}{{Schaye}}{2001}]{2001ApJ...559..507S}
{Schaye} J.,  2001, \mn@doi [\apj] {10.1086/322421}, \href
  {http://adsabs.harvard.edu/abs/2001ApJ...559..507S} {559, 507}

\bibitem[\protect\citeauthoryear{{Shapiro} \& {Giroux}}{{Shapiro} \&
  {Giroux}}{1987}]{1987ApJ...321L.107S}
{Shapiro} P.~R.,  {Giroux} M.~L.,  1987, \mn@doi [\apjl] {10.1086/185015},
  \href {http://adsabs.harvard.edu/abs/1987ApJ...321L.107S} {321, L107}

\bibitem[\protect\citeauthoryear{{Sharma}, {Theuns}, {Frenk}, {Bower}, {Crain},
  {Schaller}  \& {Schaye}}{{Sharma} et~al.}{2017}]{2017MNRAS.468.2176S}
{Sharma} M.,  {Theuns} T.,  {Frenk} C.,  {Bower} R.~G.,  {Crain} R.~A.,
  {Schaller} M.,   {Schaye} J.,  2017, \mn@doi [\mnras] {10.1093/mnras/stx578},
  \href {http://adsabs.harvard.edu/abs/2017MNRAS.468.2176S} {468, 2176}

\bibitem[\protect\citeauthoryear{{Shukla}, {Mellema}, {Iliev}  \&
  {Shapiro}}{{Shukla} et~al.}{2016}]{2016MNRAS.458..135S}
{Shukla} H.,  {Mellema} G.,  {Iliev} I.~T.,   {Shapiro} P.~R.,  2016, \mn@doi
  [\mnras] {10.1093/mnras/stw249}, \href
  {http://ukads.nottingham.ac.uk/abs/2016MNRAS.458..135S} {458, 135}

\bibitem[\protect\citeauthoryear{{Sobacchi} \& {Mesinger}}{{Sobacchi} \&
  {Mesinger}}{2014}]{2014MNRAS.440.1662S}
{Sobacchi} E.,  {Mesinger} A.,  2014, \mn@doi [\mnras] {10.1093/mnras/stu377},
  \href {http://ukads.nottingham.ac.uk/abs/2014MNRAS.440.1662S} {440, 1662}

\bibitem[\protect\citeauthoryear{{Teyssier}}{{Teyssier}}{2002}]{2002A&A...385..337T}
{Teyssier} R.,  2002, \mn@doi [\aap] {10.1051/0004-6361:20011817}, \href
  {http://cdsads.u-strasbg.fr/abs/2002A\%26A...385..337T} {385, 337}

\bibitem[\protect\citeauthoryear{{Totani} et~al.,}{{Totani}
  et~al.}{2014}]{2014PASJ...66...63T}
{Totani} T.,  et~al., 2014, \mn@doi [\pasj] {10.1093/pasj/psu032}, \href
  {http://cdsads.u-strasbg.fr/abs/2014PASJ...66...63T} {66, 63}

\bibitem[\protect\citeauthoryear{{Watkinson}, {Mesinger}, {Pritchard}  \&
  {Sobacchi}}{{Watkinson} et~al.}{2015}]{2015MNRAS.449.3202W}
{Watkinson} C.~A.,  {Mesinger} A.,  {Pritchard} J.~R.,   {Sobacchi} E.,  2015,
  \mn@doi [\mnras] {10.1093/mnras/stv499}, \href
  {http://adsabs.harvard.edu/abs/2015MNRAS.449.3202W} {449, 3202}

\bibitem[\protect\citeauthoryear{{Worseck} et~al.,}{{Worseck}
  et~al.}{2014}]{2014MNRAS.445.1745W}
{Worseck} G.,  et~al., 2014, \mn@doi [\mnras] {10.1093/mnras/stu1827}, \href
  {http://adsabs.harvard.edu/abs/2014MNRAS.445.1745W} {445, 1745}

\bibitem[\protect\citeauthoryear{{Wyithe} \& {Bolton}}{{Wyithe} \&
  {Bolton}}{2011}]{2011MNRAS.412.1926W}
{Wyithe} J.~S.~B.,  {Bolton} J.~S.,  2011, \mn@doi [\mnras]
  {10.1111/j.1365-2966.2010.18030.x}, \href
  {http://adsabs.harvard.edu/abs/2011MNRAS.412.1926W} {412, 1926}

\makeatother
\end{thebibliography}

\appendix

\section{Parameter values}
\label{sec:param_values}

In Table~\ref{tab:samples}, we give the best-fit values of the
parameters in Equation~(\ref{fit_Rahm}) for our simulations, averaged
over the five simulations presented in this paper.

\begin{table*}
  \caption{Best-fit values of the parameters in
    Equation~\ref{fit_Rahm} for our simulations, averaged over the
    five simulations presented in this paper.  They describe the
    average of the curves in Figure~\ref{fig:paramEvolAllSims}.  Also
    shown is the average photoionization rate in ionized regions.
    Densities are in physical units.}
  \label{tab:samples}
  \begin{tabular}{SSSSSSS}
    \hline
    $z$ & {$\Gamma_\mathrm{HI}$} & {$n_0$} & {$\alpha_1$} & {$\alpha_2$} & {$\beta$} & {$f$} \\
        & {($10^{-12}$~s$^{-1}$)} & {(cm$^{-3}$)} & & & & \\
    \hline
    {$3.0$} & {$1.13$} & {$0.0090$}  & {$-1.12$} & {$-1.65$} & {$5.32$} & {$0.018$} \\
    {$4.0$} & {$1.05$} & {$0.0093$}  & {$-0.95$} & {$-1.50$} & {$5.87$} & {$0.015$} \\
    {$5.0$} & {$0.90$} & {$0.0103$}  & {$-1.29$} & {$-1.60$} & {$5.06$} & {$0.024$} \\
    {$6.0$} & {$0.34$} & {$0.0070$}  & {$-0.94$} & {$-1.51$} & {$6.11$} & {$0.029$} \\
    {$7.0$} & {$0.07$} & {$0.0027$}  & {$-0.86$} & {$-1.27$} & {$7.08$} & {$0.041$} \\
    {$8.0$} & {$0.10$} & {$0.0040$}  & {$-0.74$} & {$-1.40$} & {$7.12$} & {$0.041$} \\
    {$9.0$} & {$0.14$} & {$0.0046$}  & {$-0.64$} & {$-1.21$} & {$9.99$} & {$0.029$} \\
    {$10.0$} & {$0.20$} & {$0.0047$} & {$-0.39$} & {$-0.86$} & {$12.94$} & {$0.006$} \\
    \hline
  \end{tabular}
\end{table*}

{

\section{Robustness of selection of self-shielded systems}
\label{new_method_detection_ssh}

In order to test the robustness of our Gaussian filtering technique to
isolate self-shielded regions, we test here another method to perform
this filtering.  This technique aims at isolating cells that are in
ionization equilibrium.  We do this by keeping only cells with
$\mathrm{log_{10}(R)<0.01}$, with R the ratio that gives the number of
ionized atoms per second over the number of recombined atoms per
second.  R is defined as follows:

\begin{equation}
 \mathrm{R = \frac{\Gamma \, x_{HI}} {\alpha_A \, n_H \, (1-x_{HI})^2} }
\end{equation}

Then, we exlude cells that are not yet ionized (i.e. still in neutral
region with a photoionization rate $\mathrm{\Gamma = 0}$) by keeping
only cells with $\mathrm{\Gamma \ge 10^{-17}}$.  For cells respecting
these two criteria, we compute the distribution of $\mathrm{\Gamma}$
as a function of $\mathrm{n_H}$.  Figure~\ref{newmethod} compares this
distribution with the one obtained with the Gaussian filtering method
used all over the paper.  We clearly see that the fits of the
distribution with Equation~(\ref{fit_Rahm}) obtained with the two
methods are almost indistinguishable, suggesting that the method used
all over the paper is robust.

\begin{figure*}
   \begin{center}
      \includegraphics[width=\textwidth]{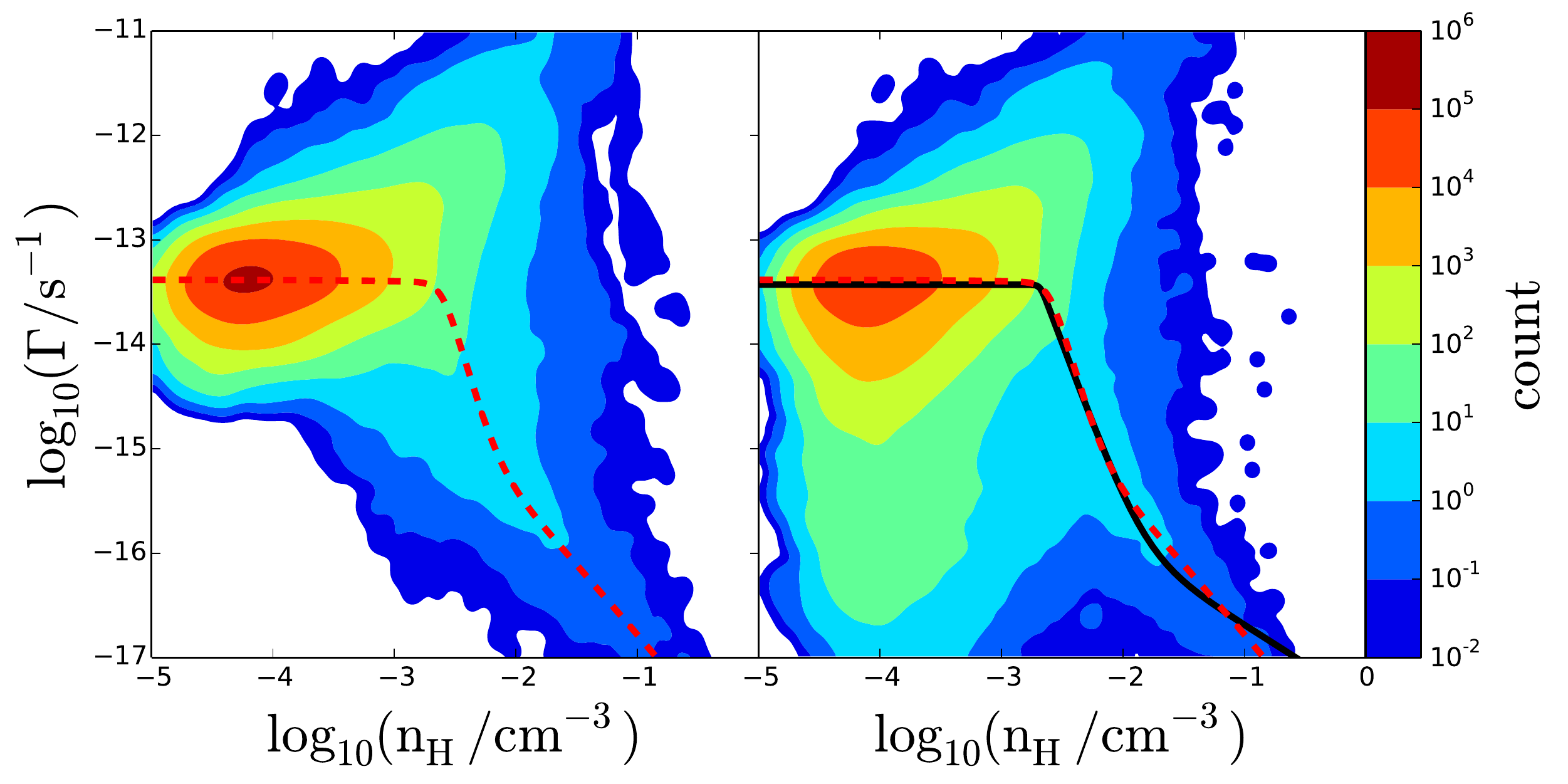}
  \caption{Comparison of two different methods of identifying ionized
    regions in our radiative transfer simulations. This is same
    simulation as in Figure~\ref{fig:fid}. Left panel shows the
    photoionization rate distribution in ionized regions identified
    using the method presented in the main text. The red dashed curve
    shows a fit using Equation~(\ref{fit_Rahm}); this curve is
    reproduced in the right panel for comparison. The left panel is
    identical to the z = 7 panel of Figure~\ref{fig:fid}. The right
    panel shows the photoionization rate distribution in ionized
    regions identified using our new selection method. In this method,
    ionized regions are defined as regions with non-zero
    photoionization rate that are in ionization equilibrium.  Black
    curve in this panel shows a fit using our
    Equation~(\ref{fit_Rahm}). The red and black curves in the right
    panel are virtually identical, which validates the selection
    method used in this paper.}
    \label{newmethod}
  \end{center}
\end{figure*}

\bsp
\label{lastpage}
\end{document}